\documentclass[aps,pre,showpacs,showkeys,amsmath,amsfonts,amssymb,notitlepage,superscriptaddress,nofootinbib]{revtex4-1} 
\usepackage[dvips]{graphicx}
\usepackage[applemac]{inputenc}
\usepackage{bbm}
\usepackage{xcolor}
\usepackage[colorlinks,linkcolor=blue,anchorcolor=blue,citecolor=blue,filecolor=blue,menucolor=blue,urlcolor=blue]{hyperref}
\usepackage{breakurl}

\newcommand{\beq}{\begin{equation}} 
\newcommand{\eeq}{\end{equation}}
\newcommand{\bea}{\begin{eqnarray}} 
\newcommand{\eea}{\end{eqnarray}}

\input epsf 
 
\begin{document} 
 
\title{Violent relaxation in the Hamiltonian Mean Field model: \\
I. Cold collapse and effective dissipation} 

\author{Guido Giachetti} 
\altaffiliation[Current address: ]{SISSA, via Bonomea 265, I-34136 Trieste, Italy}
\email{ggiachet@sissa.it} 
\affiliation{Dipartimento di Fisica e Astronomia, Università di Firenze, via G.\ Sansone 1, I-50019 Sesto Fiorentino (FI), Italy}  
\author{Lapo Casetti}
\email{lapo.casetti@unifi.it} 
\affiliation{Dipartimento di Fisica e Astronomia, Università di Firenze, via G.\ Sansone 1, I-50019 Sesto Fiorentino (FI), Italy}  
\affiliation{Istituto Nazionale di Fisica Nucleare (INFN), Sezione di Firenze, via G.\ Sansone 1, I-50019 Sesto Fiorentino (FI), Italy}  
\affiliation{INAF-Osservatorio Astrofisico di Arcetri, Largo E.\ Fermi 5, I-50125 Firenze, Italy}  

\date{\today} 
 
\begin{abstract} 
In $N$-body systems with long-range interactions mean-field effects dominate over binary interactions (collisions), so that relaxation to thermal equilibrium 
occurs on time scales that grow with $N$, diverging in the $N\to\infty$ limit. However, a much faster and completely non-collisional relaxation process, referred to as violent relaxation, sets in when starting from generic initial conditions: collective oscillations (referred to as virial oscillations) develop and damp out on timescales not depending on the system's size. After the damping of such oscillations the system is found in a quasi-stationary state that may be very far from a thermal one, and that survives until the slow relaxation driven by two-body interactions becomes effective, that is, virtually forever when the system is very large. During violent relaxation the distribution function obeys the collisionless Boltzmann (or Vlasov) equation, that, being invariant under time reversal, does not ``naturally'' describe a relaxation process. Indeed, the dynamics is moved to smaller and smaller scales in phase space as time goes on, so that observables that do not depend on small-scale details appear as relaxed after a short time. 

Here we propose an approximation scheme to describe the collisionless relaxation process, based on the introduction of suitable moments of the distribution function, and apply it to a simple toy model, the Hamiltonian Mean Field (HMF) model. To the leading order, virial oscillations are equivalent to the motion of a particle in a one-dimensional potential. Inserting higher-order contributions in an effective way, inspired by the Caldeira-Leggett model of quantum dissipation, we derive a dissipative equation describing the damping of the oscillations, including a renormalization of the effective potential and yielding predictions for collective properties of the system after the damping in very good agreement with numerical simulations. Here we restrict ourselves to ``cold'' initial conditions, i.e., where the velocities of all the particles are set to zero: generic initial conditions will be considered in a forthcoming paper. 
\end{abstract} 
 
 
 
\maketitle 

\section{Introduction} 
\label{sec_intro}

Systems with long-range interactions, i.e., interactions that decay with the distance $r$ between the interacting bodies slower than $r^{-d}$, where $d$ is the dimension of space, exhibit peculiar features. A striking peculiarity of these systems is that relaxation to thermal equilibrium occurs on time scales $\tau_{\text{rel}}$ that grow with the number $N$ of constituents of the system, diverging in the $N\to\infty$ limit \cite{CampaEtAl:book,CampaEtAl:physrep}. Thus, in practice, a many-body long-range-interacting system never reaches equilibrium and is typically found in a non-thermal state. The paradigmatic example of a system with long-range interactions that can be found in nature is a self-gravitating system where\footnote{For self-gravitating systems a proper thermal equilibrium state in the usual sense does not exist, at least in three dimensions, because gravity is non-confining so that a Maxwellian velocity distribution would lead to the evaporation of the system, unless the latter has infinite mass. The relaxation time here is the time scale over which binary encounters induce a loss of memory of the initial conditions and the Boltzmann entropy grows. Gravity in spaces with dimension less than three (where a thermal equilibrium state is well defined) also has relaxation times diverging with $N$.} $\tau_{\text{rel}} \propto N/\ln N$: for a typical elliptical galaxy the relaxation time is of the order of $10^{17}$ years, i.e., several orders of magnitude larger than the age of the Universe \cite{BinneyTremaine:book}. Other examples of long-range interactions are given by dipolar forces in three-dimensional condensed matter systems \cite{CampaEtAl:book}, unscreened electromagnetic interactions in plasmas \cite{Nicholson:book}, and effective interactions mediated by the electromagnetic field in systems of cold atoms in an optical cavity \cite{SchutzMorigi:prl2014,njp2016}. Such a striking dynamical behaviour is due to the fact that in long-range-interacting systems mean-field interactions dominate over binary encounters (collisions), so that the time evolution of the one-particle distribution function $f(\mathbf{q},\mathbf{p},t)$ is given by the collisionless Boltzmann---also referred to as Vlasov---equation for times $t < \tau_{\text{rel}}$.  Being such an equation invariant under time reversal, the Boltzmann entropy is a constant of motion and no relaxation towards thermal equilibrium can occur as long as $f$ obeys the Vlasov equation. We might therefore expect that, in the typical situation, the distribution function $f$ will not stop its evolution until the finite-$N$ collisional effects become important, bringing the system towards the thermal equilibrium state corresponding to the maximum of the Boltzmann entropy. 

However, the typical evolution is not the naively expected one: when starting from generic initial conditions, the system undergoes collective oscillations and apparently settles down after much shorter time scales, not depending on its size, remaining trapped in a non-thermal state virtually forever, when $N$ is sufficiently large. This first part of the dynamical evolution has been called ``violent relaxation'' by Donald Lynden-Bell who first proposed a statistical-mechanical approach to this phenomenon \cite{Lynden-Bell:mnras1967}. The prototypical example is the collapse of an isolated cloud of self-gravitating particles in three dimensions, where starting from the pioneering numerical works by Hénon \cite{Henon:1964AnnAp1964} and van Albada \cite{vanAlbada:mnras1982} (see \cite{Sylos:mnras2012} and references therein for more recent results) violent relaxation has been observed in the form of strong collective oscillations that damp out on a short time scale, of the order of a few dynamical times\footnote{The dynamical time $\tau_D$ of a self-gravitating system is defined as $\tau_D \approx \sqrt{\pi R^3/(GM)}$, where $R$ is the size if the system, or $\tau_D \approx \sqrt{1/(G\varrho)}$ where $\varrho$ is the average mass density, and is an estimate of the crossing time, i.e., the time needed for a typical particle to cross the entire system.}, leaving the system in a non-thermal but virialized stationary state. The collective oscillations during violent relaxation of a self-gravitating system are usually referred to as ``virial oscillations'', since the virial ratio $2K/U$, where $K$ is the total kinetic energy and $U$ is the total gravitational potential energy, oscillates around, and eventually sets to, the stationary value $2K/U = -1$. In the last decades it has become clear that violent relaxation is not a peculiarity of three-dimensional self-gravitating systems: for instance, such a phenomenon has been observed in two-dimensional self-gravitating systems, when they collapse from a dilute state as well as when an initially thermal state is strongly perturbed \cite{mnras2018}, in mean-field toy models, again when an initially thermal state is strongly perturbed \cite{prerap2015}, and in semiclassical models of cold atoms in an optical cavity, after a quench from a thermal equilibrium state \cite{njp2016}. It is now clear that virial oscillations and violent relaxation are universal properties of long-range-interacting systems, occurring with essentially the same features as in the gravitational collapse case \cite{CampaEtAl:book}. After the damping of the virial oscillations the system is found in a non-thermal state that is usually referred to as a quasi-stationary state (QSS), to indicate that such a state is not a true equilibrium of the system, being not stable under the action of collisions; in the $N \to \infty$ limit, however, quasi-stationary states become true stationary states of the Vlasov equation. Given that long-range-interacting systems spend virtually an infinite amount of time in quasi-stationary states, their prediction and characterization becomes maybe the most relevant task concerning this class of physical systems. However, we are still not able to predict the actual state in which the system is going to settle down after violent relaxation as, for example, the final state reached after a gravitational collapse, in the self-gravitating case. Indeed, to make a physically relevant example, elliptical galaxies are thought to be in a quasi-stationary state---also on purely observational grounds, the fact that most elliptical galaxies do share the same kind of luminosity profiles is an indication that some non-collisional relaxation process has occurred---but we are not able to predict such a state from first principles, given the initial conditions, and it is not even clear actually how pronounced the dependence on the initial conditions is \cite{BinneyTremaine:book,Bertin:book}. 

Theoretical approaches to violent relaxation started with Lynden-Bell's theory \cite{Lynden-Bell:mnras1967}, motivated by the attempt to predict the density profile of elliptical galaxies by means of a statistical-mechanical-like approach; however, it appears that Lynden-Bell's theory only works well when virial oscillations are somehow suppressed \cite{LevinEtAlphysrep:2014,Sylos:mnras2012}. Despite many advances since then (see e.g.\ \cite{CampaEtAl:book,LevinEtAlphysrep:2014} for reviews and \cite{LeonciniVanDenBergFanelli:epl2009,CampaChavanis:jstat2010,ChavanisCampa:epjb2010,AssllaniEtAl:pre2012,BenettiEtAl:prl2014} for applications to particular systems), a satisfactory theory of violent relaxation is still lacking for any long-range-interacting systems, even the simpler ones, not only for galaxies. The phenomenology of virial oscillations and violent relaxation suggests that the mechanism behind such a collisionless relaxation is similar to Landau damping in plasmas, where the dynamics is reversible but the collective oscillations lose their energy towards the motions of the particles via resonant interaction (see e.g.\ the discussion in \cite{Kandrup:apj1998}). However, this kind of Landau damping would be fully nonlinear and would occur in non-homogeneous states; at variance with the linear and homogeneous case that admits an exact solution in terms of uncoupled normal modes, the nonlinear and non-homogeneous case presents technical challenges that have not been overcome yet, although progress has been made in particular cases  \cite{BarreOlivettiYamaguchi:jstat2010,BarreOlivettiYamaguchi:jphysa2011}. It is worth mentioning that a phase-space interpretation of collisionless relaxation is possible (see e.g.\ \cite{TremaineHenonLynden-Bell:mnras1986}, or \cite{MohoutVillani:actamath2011} for a more mathematically-oriented discussion), and would be useful in the following. As time progresses, the Vlasov evolution splits the distribution function $f$ in finer and finer filaments that fold on themselves. As a consequence the dynamics relocates on smaller and smaller scales in phase space: then, rigorously speaking, the evolution in phase space does not admit any asymptotic state. However, any coarse-grained distribution function $\tilde{f}$ will not be affected by this fine dynamics and will eventually stop its evolution, so that the coarse graining actually introduces an effective time arrow in the system. Despite the dynamics would never really stop, any physically meaningful observable related to not too small scales will settle to a pseudo-equilibrium value.

In the present paper we aim at giving a contribution to the understanding of this problem. In order to reduce the complexity, we shall restrict ourselves to a very simplified toy model, the Hamiltonian Mean Field (HMF) model (see Sec.\ \ref{sec_HMF}), that, despite being one of the simplest models of a long-range-interacting systems, does exhibit all the above mentioned phenomenology. Moreover, we shall restrict ourselves to ``cold'' initial conditions, i.e., where the velocities of all the particles are set to zero. We shall introduce a hierarchy of moments of the distribution function, in order to study the dynamics of the lowest-order moments by taking into account the contribution of the higher-order ones in an effective way, in close analogy to the Caldeira-Leggett model of quantum dissipation \cite{CaldeiraLeggett:prl1981,CaldeiraLeggett:physa1983}: as we shall see, this will allow us to write an effective equation of motion for the low-order moments where the contribution of the other moments results in an explicit dissipation as well as a renormalization of the low-order dynamics. In a forthcoming paper \cite{violent_rel_II} we shall generalize the approach introduced here to generic initial conditions.

The paper is organized as follows: in Sec.\ \ref{sec_HMF} we introduce the HMF model, in Sec.\ \ref{sec_cold} we introduce our theoretical approach, apply it to the cold collapse of the HMF model and compare its predictions with numerical simulations, and in Sec.\ \ref{sec_conclusions} we draw our conclusions and discuss open points and future developments. Two appendices are devoted to technical and/or side aspects.   

\section{Hamiltonian Mean Field model}
\label{sec_HMF}

The Hamiltonian Mean Field (HMF) model is a toy model that has become paradigmatic for the study of equilibrium as well as nonequilibrium properties of systems with long-range interactions. According to Chavanis and Campa \cite{ChavanisCampa:epjb2010}, it was introduced by Messer and Spohn \cite{MesserSpohn:jsp1982}---who called it the ``cosine model''---after a suggestion by Battle \cite{Battle:cmp1977}; the model was then made popular by Antoni and Ruffo,  who also introduced the name and the acronym HMF, in a seminal paper \cite{AntoniRuffo:pre1995}. Many works have been devoted to the study of the HMF model since then: a review of the main results can be found in \cite{CampaEtAl:book,CampaEtAl:physrep}. 
The HMF model is a classical system with $N$ degrees of freedom, defined by the Hamiltonian\footnote{Note that our Hamiltonian \eqref{eq:H_HMF} differs---by an additive constant---from the one used in many works on the HMF model where the interaction energy is chosen such as to be always larger than zero.}
\beq
\mathcal{H} = \frac{1}{2}\sum_{i = 1}^N p_i^2 - \frac{1}{2N}\sum_{i,j = 1}^N \cos\left(\vartheta_i -\vartheta_j\right)~,
\label{eq:H_HMF} 
\eeq 
where $\vartheta_i \in [-\pi,\pi]$ are angular coordinates, $p_i$ are their conjugated momenta and we have chosen natural units such that the coupling constant and the inertia of each degree of freedom are equal to unity. We note that we chose the coupling constant to be positive, so that the interaction is attractive (but a repulsive version of the HMF has been considered too, see e.g.\ \cite{AntoniRuffo:pre1995,CampaEtAl:book}) and the Kac scaling has been used, such that the Hamiltonian \eqref{eq:H_HMF} is extensive, although clearly non additive.

The HMF model admits (at least) two different interpretations. It can be seen either as a fully connected system of planar ($XY$) classical spins, each with a finite inertia and parametrized by one of the angles $\vartheta_i$, interacting via a classical, ferromagnetic Heisenberg exchange interaction\footnote{The acronym HMF can also stand for Heisenberg Mean Field.}, or as a system of classical particles moving on a circle, whose center coincides with the origin of the reference frame and where the angles $\vartheta_i$ are the angular polar coordinates of each particle, interacting via the potential
\beq
U\left(\vartheta_1,\ldots,\vartheta_N \right) = \frac{1}{2N}\sum_{i,j = 1}^N v\left(\vartheta_i -\vartheta_j\right)= - \frac{1}{2N}\sum_{i,j = 1}^N \cos\left(\vartheta_i -\vartheta_j\right)~.
\label{eq:Uparticles}
\eeq
Although the particle interpretation is the most natural in our context, the terminology commonly used for the HMF model stems from the magnetic interpretation. Introducing the magnetization vector $\mathbf{m} = \left(m_x,m_y\right)$ such that
\begin{subequations}
\begin{align}
m_x & = \frac{1}{N} \sum_{i=1}^N \cos\vartheta_i\,,
\label{eq:mx}\\
m_y & = \frac{1}{N} \sum_{i=1}^N \sin\vartheta_i\,,
\label{eq:my}
\end{align}  
\label{eq:m}
\end{subequations}
the interaction energy \eqref{eq:Uparticles} becomes
\beq
U = - \frac{1}{2N}\sum_{i,j = 1}^N \cos\left(\vartheta_i -\vartheta_j\right)= -\frac{1}{2}m^2~,
\eeq
where $m = \left| \mathbf{m} \right| = \sqrt{m_x^2 + m_y^2}$. The vector $\mathbf{m}$ is indeed the (instantaneous) magnetization of the system \eqref{eq:H_HMF}; in the particle interpretation, it is a vector pointing to the direction of the maximum particle density on the circle and whose modulus is related to the degree of clustering of the particles: $m=0$ corresponds to a uniform density and $m=1$ to a completely collapsed (delta-like) distribution of particles\footnote{The square modulus $m^2$ of the magnetization plays an analogous  r\^{o}le to the reciprocal of the gravitational radius in a self-gravitating system.}. The Hamilton equations of motion derived from \eqref{eq:H_HMF} can be written as
\begin{subequations}
\begin{align}
\dot\vartheta_i & = p_i \,,
\label{eq:ham_theta}\\
\dot p_i & = - m_x \sin\vartheta_i - m_y\cos\vartheta_i \,,
\label{eq:ham_p}
\end{align}  
\label{eq:ham}
\end{subequations}
where it is apparent that the coupling between the degrees of freedom comes only from the magnetization, whence the ``mean field'' in the name of the system.

In the limit $N\to\infty$ we can introduce the single-particle distribution function $f(\vartheta,p,t)$ and replace the sums $\frac{1}{N}\sum_{i = 1}^N$ with the phase space averages weighted by $f$, i.e., the integrals $\int_{-\infty}^{+\infty}dp \int_{-\pi}^{\pi} d \vartheta\, f$. The magnetization components become functionals of $f$,
\begin{subequations}
\begin{align}
m_x[f] & =  \int_{-\infty}^{+\infty}dp \int_{-\pi}^{\pi} d \vartheta \, f(\vartheta,p,t) \cos\vartheta = \langle \cos\vartheta \rangle\,,
\label{eq:mxcont}\\
m_y[f] & = \int_{-\infty}^{+\infty}dp \int_{-\pi}^{\pi} d \vartheta \, f(\vartheta,p,t) \sin\vartheta = \langle \sin\vartheta \rangle\,,
\label{eq:mycont}
\end{align}  
\label{eq:mcont}
\end{subequations}
the mean-field potential is given by
\beq
U[f](\vartheta) = - \int_{-\infty}^{+\infty}dp' \int_{-\pi}^{\pi} d \vartheta' \, f(\vartheta',p',t) \cos\left(\vartheta - \vartheta' \right) = - \cos\vartheta \langle \cos\vartheta' \rangle - \sin\vartheta \langle \sin\vartheta' \rangle = -m_x[f] \cos\vartheta - m_y[f] \sin\vartheta\,,
\eeq
so that the distribution function evolves in time according to the Vlasov equation
\beq
\frac{\partial f}{\partial t} + p\frac{\partial f}{\partial \vartheta} - \left(m_x[f] \sin\vartheta - m_y[f]\cos\vartheta \right)\frac{\partial f}{\partial p} = 0\,.
\label{eq:Vlasov_HMF}
\eeq
It is worth noticing that the mean-field interaction of the HMF model can be seen as the lowest-order expansion of a generic mean-field interaction for particles on a circle. Indeed, on a circle any interaction must be periodic, so that we can expand it in a Fourier series,
\beq
U[f](\vartheta) = A_0[f] + \sum_{n=1}^\infty A_n[f] \cos \left( n\vartheta \right) + \sum_{n=1}^\infty B_n[f] \sin \left( n\vartheta \right) \,,
\eeq
and neglecting the constant term $A_0$ by redefining the zero level of the interaction and keeping only the lowest order in the expansion, we get
\beq
U[f](\vartheta) = A_1[f] \cos \vartheta  + B_1[f] \sin \vartheta \,.
\eeq  
Requiring rotational invariance on the circle, the only possible preferred direction is that specified by the mean position of the particles, implying
\begin{subequations}
\begin{align}
A_1[f] & =  \mu\langle \cos\vartheta \rangle = \mu m_x \,, \\
B_1[f] & =  \mu\langle \sin\vartheta \rangle = \mu m_y \,. 
\end{align}  
\label{eq:mu}
\end{subequations}
Being the mean-field potential defined as
\beq
U[f](\vartheta) = \int_{-\infty}^{+\infty}dp' \int_{-\pi}^{\pi} d \vartheta' \, f(\vartheta',p')\, v(\vartheta - \vartheta')\,,
\eeq
it is linear in $f$, so that the quantity $\mu$ in Eqs.\ \eqref{eq:mu} cannot depend of $f$ and is therefore a scalar constant that can be absorbed in the choice of units and set equal to $-1$, since it has to be negative if we want the potential to be attractive. We finally get
\beq
U[f](\vartheta) = - m_x[f]\cos\vartheta - m_y[f]\sin\vartheta\,,
\eeq  
i.e., the mean-field potential of the HMF model, that can thus be regarded as a set of particles interacting with the lowest Fourier modes of a generic attractive collective interaction on a circle. Indeed, it has been shown that the HMF interaction stems from a softened gravitational interaction between particles constrained on a circle, in the limit of an infinitely large softening length \cite{TatekawaEtAl:pre2005}. 

From now on we shall only consider initial conditions that are symmetric around $\vartheta = 0$ and with a total momentum equal to zero, i.e., such that $f(\vartheta,p,0) = f(-\vartheta,-p,0)$. Being such invariance conserved by the equations of motion, in their discrete version \eqref{eq:ham} as well as in the continuum Vlasov limit \eqref{eq:Vlasov_HMF}, the distribution function will be such that 
\beq
f(\vartheta,p,t) = f(-\vartheta,-p,t) 
\label{eq:fsymm}
\eeq
at any time $t$; this implies not only a vanishing total momentum but also $m_y \equiv 0$, so that the magnetization will be always along the $\vartheta = 0$ axis and $m \equiv m_x$. The Vlasov equation then becomes
\beq
\frac{\partial f}{\partial t} + p\frac{\partial f}{\partial \vartheta} - m[f] \sin\vartheta \frac{\partial f}{\partial p} = 0\,.
\label{eq:Vlasov_HMF_symm}
\eeq
The periodic boundary conditions imply that $f(\pi,p) = f(-\pi,p)$, so that the single-particle phase space has the topology of a cylinder; this, together with the condition \eqref{eq:fsymm}, means that we can also see it as the sheet $\vartheta\in[0,\pi]$ in the $(\vartheta,p)$ Euclidean plane, with two perfectly elastic walls in $\vartheta = 0$ and $\vartheta = \pi$. 

As a final remark for this Section, since we are mainly concerned with the HMF model as a toy model of a self-gravitating system, it is worth mentioning that the HMF model has been considered in an astrophysical context in \cite{Pichon:thesis} and, more recently, in \cite{FouvryBarOr:mnras2018}.

\section{Cold collapse in the HMF model}
\label{sec_cold}

\begin{figure}
\includegraphics[width = 0.7\textwidth]{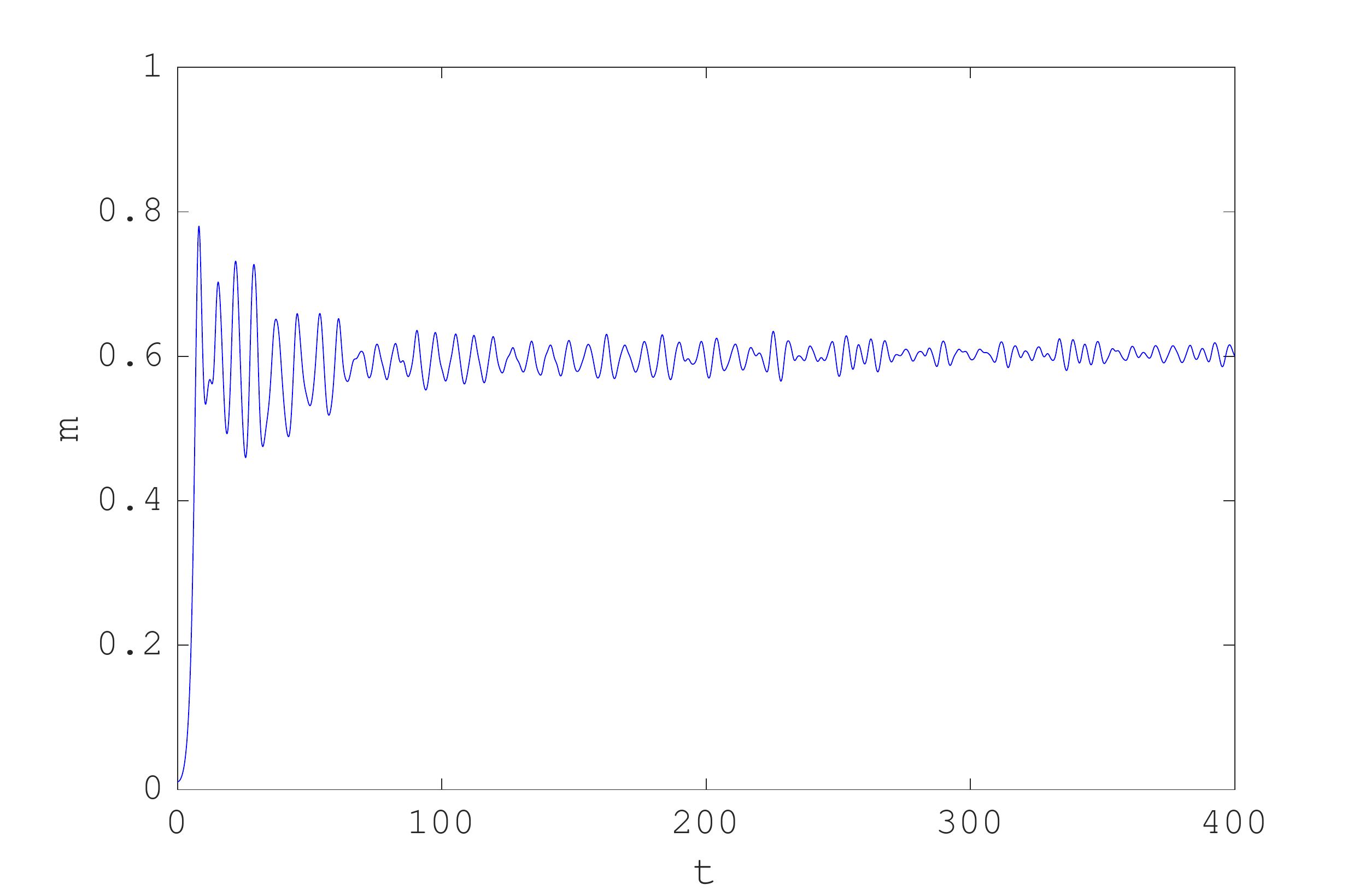}
\caption{Magnetization $m$ as a function of time for an HMF model with $N = 2\times 10^6$. Cold initial conditions ($p_i = 0~\forall i$) with a uniform distribution of particles in $[-\pi,\pi]$ and a Gaussian overdensity made of a fraction of $10^{-2}$ of the total number of particles, such that $m_0 \approx 0.011$.}
\label{fig:coldcollapse}  
\end{figure}

Let us now consider a cold initial distribution function, i.e., such that all the velocities are equal to zero, peaked (even very weakly) around $\vartheta = 0$ (so that $m(0) = m_0 > 0$, with $m_0$ possibly very small). As one can easily guess, the attractive nature of the potential will immediately lead to a collapse that, in turn, results in the growth of $m$ and then in the loss of potential energy that is turned into kinetic energy. After reaching the minimum of the collective potential, particles will start to turn kinetic energy back into potential energy as the system bounces back to a less clustered state. We thus expect a behaviour that reminds of a Jeans-like instability, where an initial collapse is followed by collective oscillations, analogous to the virial oscillations observed in a self-gravitating system. Indeed, we may expect the dynamical evolution of an HMF model after a cold initial condition to be very similar to a gravitational collapse, and numerical simulations show that this is true. Moreover, simulations show that, as in the gravitational case, after a relatively short time violent relaxation sets in, the virial oscillations damp out and the system appears to relax to a quasi-stationary state, where collective quantities as $m$ are nearly constant. An example of the behaviour of the magnetization $m(t)$ of an HMF model with $N = 2\times 10^6$ particles in a cold collapse starting with very small magnetization ($m_0 \approx 10^{-2}$) is shown in Fig.\ \ref{fig:coldcollapse}. The simulation whose results are shown in Fig.\ \ref{fig:coldcollapse} as well as all the other simulations of the HMF model whose results are shown in the following have been conducted by integrating Eqs.\ \eqref{eq:ham} by means of a third-order bilateral symplectic algorithm \cite{physscr1995}.

\subsection{Moments of inertia of the distribution function}
\label{subsec_moments}

Let us now introduce our theoretical approach to virial oscillations and collisionless relaxation in a cold collapse of the HMF model. First of all, let us observe that all the states of the system are expected to become rather tightly collapsed, even after starting with small magnetization, as shown in Fig.\ \ref{fig:coldcollapse}. We can thus replace the $\sin\vartheta$ term in the Vlasov equation \eqref{eq:Vlasov_HMF_symm} with its Taylor expansion up to a finite order $2J+1$, so that the time evolution of $f$ is given by
\beq
\frac{\partial f}{\partial t} + p\frac{\partial f}{\partial \vartheta} - m[f] \left[ \sum_{j=0}^J \frac{(-1)^j}{(2j + 1)!} \vartheta^{2j+1}\right] \frac{\partial f}{\partial p} = 0\,.
\label{eq:Vlasov_HMF_taylor}
\eeq
For finite and not very large values of $J$ this approximation will work better when the state is tightly collapsed, as noted before: in spite of that, for nearly homogeneous states $m \approx 0$ and the depth of the potential well $-m\cos\vartheta$ is small, so that the actual form of the potential is quite irrelevant (as long as the minimum remains in $\vartheta = 0$). 

\begin{figure}
\includegraphics[width = 0.7\textwidth]{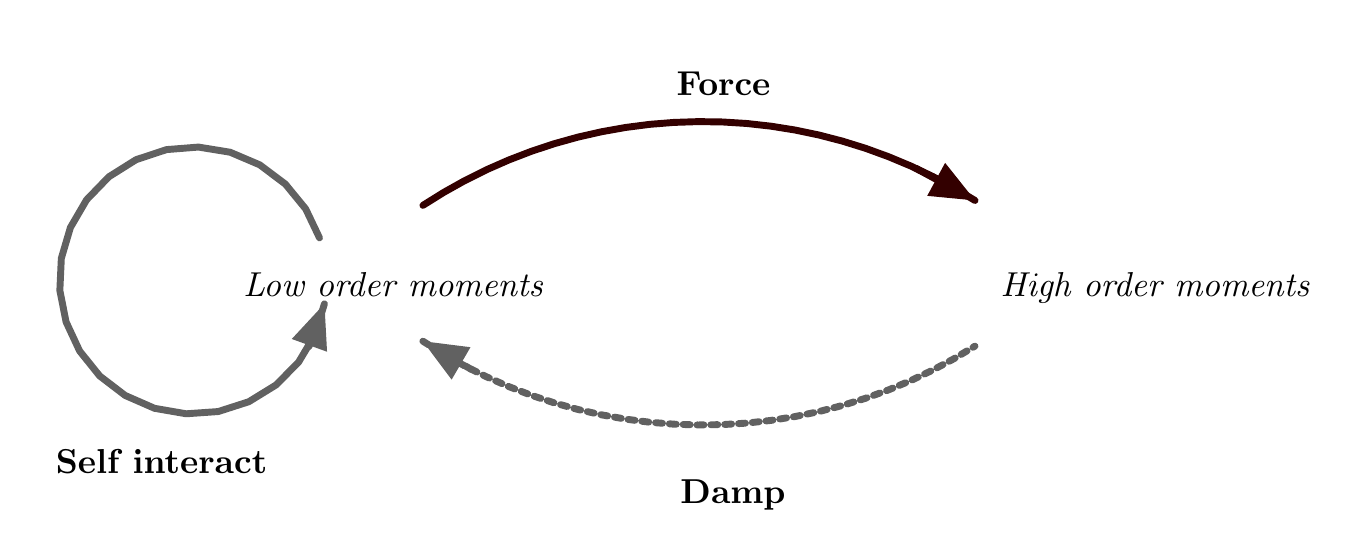}
\caption{Schematic representation of the hierarchy of interactions between the moments $I_{k,n}$ as given by Eqs.\ \eqref{eq:Idot} and \eqref{eq:mJ}.}
\label{fig:interactions}  
\end{figure}

The truncated potential from which the polynomial force given by the finite expansion in \eqref{eq:Vlasov_HMF_taylor} arises is somewhat analogous to the Villain approximation to the XY model \cite{Villain:jphys1975}, although the latter is usually truncated at the harmonic order, while we consider a generic finite order. However, such a force is discontinuous in $\vartheta = \pm \pi$ on the circle; to avoid this complication, that would make the distribution function nondifferentiable, we ``open'' the ring and insert elastic walls in $\vartheta = \pm \pi$, replacing Eq.\ \eqref{eq:Vlasov_HMF_taylor} with 
\beq
\frac{\partial f}{\partial t} + p\frac{\partial f}{\partial \vartheta} - m[f] \left[ \sum_{j=0}^J \frac{(-1)^j}{(2j + 1)!} \vartheta^{2j+1}\right] \frac{\partial f}{\partial p} +pf\left[\delta(\vartheta+\pi) -  \delta(\vartheta-\pi)\right]= 0\,.
\label{eq:Vlasov_HMF_taylor_walls}
\eeq
We now define the generalized moments of inertia of the distribution function $f$ as follows
\beq
I_{k,n}(t) = \langle \vartheta^k p^n \rangle = \int_{-\infty}^{+\infty}dp \int_{-\pi}^{\pi} d \vartheta \, f(\vartheta,p,t)\, \vartheta^k p^n \,.
\label{eq:Ikn}
\eeq
Higher order moments probe the finer grain of the distribution, so that we expect the lower order moments, blind to the fine structure, to settle before the others. To obtain an equation of motion for the moments $I_{k,n}$ we take the time derivative of the right-hand-side of Eq.\  \eqref{eq:Ikn}; by exchanging integration and differentiation, and using Eq.\ \eqref{eq:Vlasov_HMF_taylor_walls} to express $\partial_t f$, after integrations by parts and assuming that $f$ decays sufficiently fast for large $p$'s we get rid of the boundary terms and obtain a system of equations that contains only the moments, 
\beq
\dot I_{k,n} = k \, I_{k-1,n+1} - nm\sum_{j=0}^J \frac{(-1)^j}{(2j + 1)!} I_{k+2j+1,n-1}\,,
\label{eq:Idot}
\eeq
where $m$ can be expressed in terms of the moments as the following series up to order $2J+2$,
\beq
m = \sum_{j=0}^{J+1} \frac{(-1)^j}{(2j)!} I_{2j,0}\,.
\label{eq:mJ}
\eeq
The quantity $m$ introduced in Eq.\ \eqref{eq:mJ} is the truncated magnetization, and the truncation is made at an order consistent to that of the finite polynomial expansion of the $\sin\vartheta$ term in the Vlasov equation \eqref{eq:Vlasov_HMF_taylor}. Hence, Eqs.\ \eqref{eq:Idot} and \eqref{eq:mJ} yield a closed system of equations at any given order $J$: $J = 0$ involves only moments with $k+n \le 2$, $J = 1$ moments with $k+n \le 4$, and so on.  
We note that $I_{0,0}\equiv 1$ and that, due to the fact that $f$ is even in $\vartheta$ and $p$, $I_{k,n} \not = 0$ only if $k+n$ is even. Equations \eqref{eq:Idot} and \eqref{eq:mJ} clearly show the hierarchy in the interactions: lower order moments strongly interact with each other while are only weakly affected by the higher order ones, which in turn are strongly forced by the lower order moments. In particular, the dynamics of the low-order moments will describe the virial oscillations, and we expect damping to be a result of the back-action of the higher-order inertia moments, that interfere incoherently with the virial oscillations as depicted in the scheme drawn in Fig.\ \ref{fig:interactions}.
In the following we shall elaborate on this idea.

\subsection{Leading order description of virial oscillations}
\label{subsec_virial}

\begin{figure}
\includegraphics[width = 0.7\textwidth]{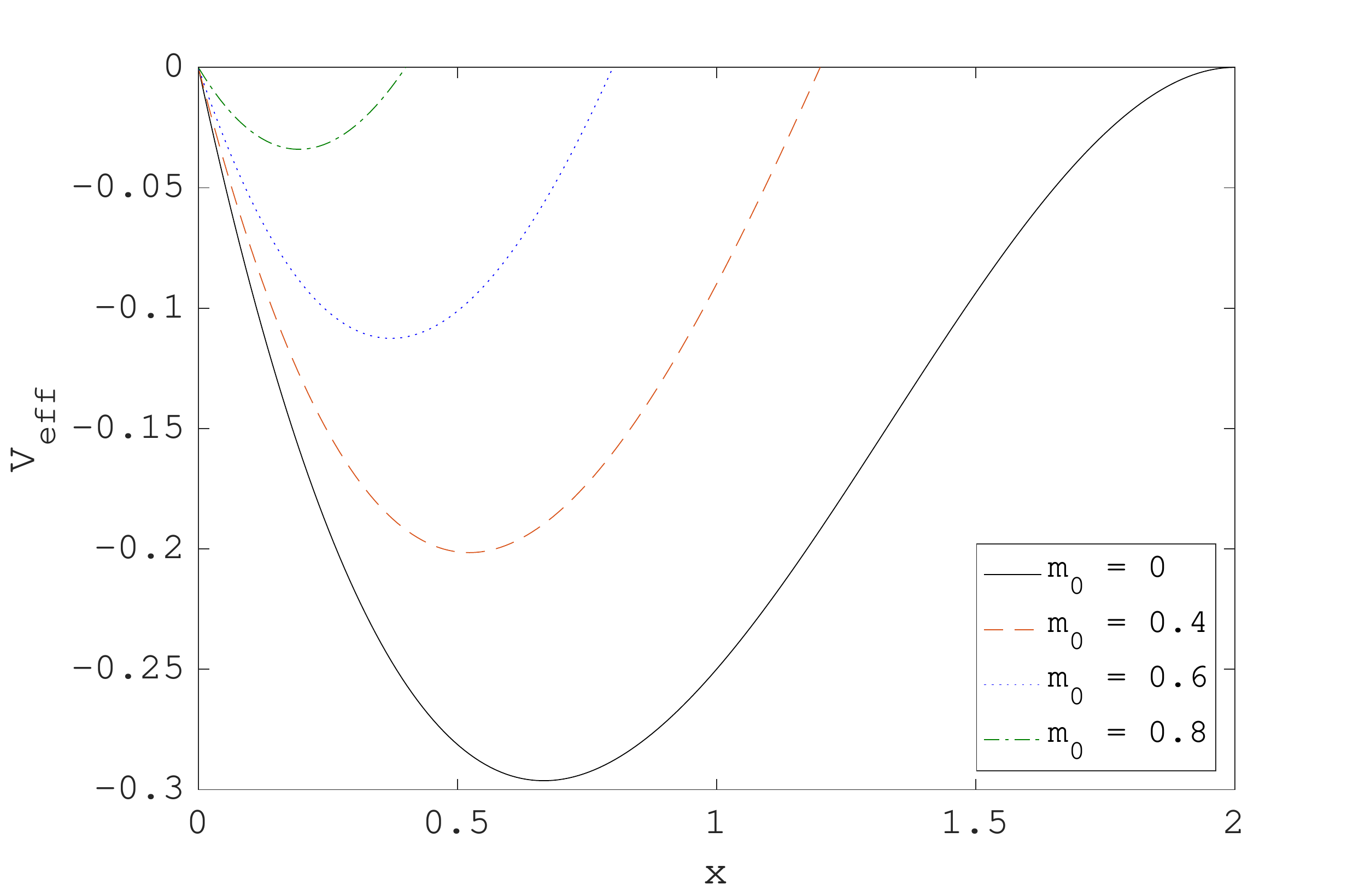}
\caption{Effective potential $V_{\text{eff}}$ given by Eq.\ \eqref{eq:veff_cold} as a function of $x$ in the case of cold collapse for some values of the initial magnetization $m_0$ (see inset).}
\label{fig:veff}  
\end{figure}

The leading order ($J = 0$,  $k+n \le 2$) in our approximation scheme involves a self-consistent harmonic approximation for the mean-field potential. The only moments to be considered are $I_{2,0} = \langle \vartheta^2 \rangle$, related to the width of the spatial distribution of the particles, $I_{0,2} = \langle p^2 \rangle$, proportional to the average kinetic energy, and $I_{1,1} = \langle \vartheta p \rangle$, i.e., the covariance between space and velocity variables. Their time evolution is given by Eqs.\ \eqref{eq:Idot} and \eqref{eq:mJ} that reduce, to the leading order, to
\begin{subequations}
\begin{align}
\dot I_{0,2} & =  2I_{1,1}\\
\dot I_{1,1} & =  I_{0,2} - m I_{2,0}\\
\dot I_{0,2} & =  -2m I_{1,1}
\end{align}
\label{eq:Idot_LO}
\end{subequations}
and
\beq
m = 1 - \frac{1}{2}I_{2,0}~.
\label{eq:mJ_LO}
\eeq  
By setting, to ease the notation, $x =  I_{0,2}$, $y = I_{1,1}$ and $z = I_{0,2}$, Eqs.\ \eqref{eq:Idot_LO} and \eqref{eq:mJ_LO} can be rewritten as
\begin{subequations}
\begin{align}
\dot x & =  2y\,, \label{eq:dotx}\\
\dot y & =  z - \left(1 - \frac{1}{2} x \right) x\,, \label{eq:doty}\\
\dot z & =  -2 \left(1 - \frac{1}{2} x \right) y\, \label{eq:dotz}.
\end{align}
\label{eq:xyz_LO}
\end{subequations}
From the above equations we get 
\beq
\dot z =  - \left(1 - \frac{1}{2} x \right) \dot x = -\frac{d}{dt}\left(1 - \frac{1}{2} x \right)^2\,,
\eeq
so that Eqs.\ \eqref{eq:xyz_LO} admit the integral of motion
\beq
\varepsilon = \frac{z}{2} - \frac{1}{2}\left(1 - \frac{1}{2} x \right)^2 = \frac{\langle p^2 \rangle}{2} - \frac{m^2}{2}~,
\label{eq:eps_LO}
\eeq
that is, the energy per particle: as it should be, $\varepsilon \ge -1/2$. Inserting Eq.\ \eqref{eq:eps_LO} into Eq.\ \eqref{eq:doty} we get
\begin{subequations}
\begin{align}
\dot x & =  2y\,, \label{eq:dotx_ham}\\
\dot y & =  2\varepsilon + 1 -2x +  \frac{3x^2}{4} \,, \label{eq:doty_ham}
\end{align}
\label{eq:xy_LO_ham}
\end{subequations}
that is, Hamilton equations of motion of a particle of mass $1/2$ moving in an effective potential $V_{\text{eff}}(x)$; indeed Eqs.\ \eqref{eq:xy_LO_ham} can be cast as
\beq
\frac{1}{2}\ddot x = -\frac{d}{dx} V_{\text{eff}}(x)\,,
\label{eq:ddotx}
\eeq
where
\beq
V_{\text{eff}}(x) = \left(2\varepsilon + 1 \right)x + x^2 - \frac{x^3}{4}~.
\label{eq:veff}
\eeq
Clearly also the energy of this motion, 
\beq
\Lambda = \frac{1}{4}\dot x^2 + V_{\text{eff}}(x) = y^2 + V_{\text{eff}}(x)\,,
\eeq 
is a constant. We note that
\beq
V_{\text{eff}} = -x\left[2\varepsilon + \left(1 - \frac{x}{2} \right)^2 \right] = -xz \,,
\eeq 
so that $\Lambda = y^2 - xz$; since the definition of the $I_{k,n}$ implies $y^2 \le xz$, one has
\beq
\Lambda = y^2 - xz \le 0\,. 
\label{eq:diseq_Lambda}
\eeq
The case of completely cold initial conditions, i.e., of vanishing initial kinetic energy, translates into the condition $z_0 = z(0) = 0$. Let us then considering the initial conditions on $y$ and $x$. As far as $x_0 = x(0)$ is concerned, we have
\beq
x_0 = 2\left( 1 - m_0\right) \in [0,2]\,,
\label{eq:x0}
\eeq 
where $m_0$ is the initial magnetization. 
Since $y^2 \le xz$, Eq.\ \eqref{eq:x0} implies $y_0 = y(0) = 0$, while Eq.\  \eqref{eq:diseq_Lambda} implies $\Lambda = 0$, so that a cold collapse corresponds to a motion with zero total energy and vanishing initial velocity (thus, with zero initial potential energy) in the effective potential \eqref{eq:veff}. The total energy is completely specified by the initial magnetization according to
\beq
\varepsilon = -\frac{1}{2}m_0^2 
\eeq
so that $\varepsilon \in \left[-\frac{1}{2}, 0\right]$ and $V_{\text{eff}}$ can be expressed  as
\beq
V_{\text{eff}}(x) = \left(1 - m_0^2\right)x + x^2 - \frac{x^3}{4}
\label{eq:veff_cold}
\eeq
and is plotted as a function of $x$ for some values of $m_0$ in Fig.\ \ref{fig:veff}. The particle starts at $x_0$ with zero velocity and, since $V_{\text{eff}}(x_0) = 0$, oscillates in the allowed region given by $V_{\text{eff}} \le 0$, that is 
\beq
x(t) \in \left[0,x_0 \right]\,.
\eeq
In the oscillations the particle always reaches $x = 0$, that is $ m = 1$. This is coherent with the expectation, and the numerical evidence, that in a cold collapse the system performs large-amplitude oscillations and always comes close to a maximally collapsed configuration. Initially larger magnetizations (corresponding to smaller $x_0$) give rise to smaller-amplitude oscillations, being the latter always bounded between $0$ and $x_0$, again a feature that is observed in simulations.  
When $m_0 = 1$ the potential has a minimum in $x_0 = 0$ and no evolution is possible, since the allowed region for $x$ shrinks to the point $x_0$ itself: this makes sense because the whole system is in the bottom of the $N$-body potential well. Also the initial position $x_0 = 2$, corresponding to $m_0 = 0$, is a stationary state of the potential, but it is a  maximum, so that any $m_0$ slightly larger than zero would lead to a collapse. As expected, this is an unstable equilibrium, that corresponds to the instability of a cold uniform distribution. Hence, the leading-order description we have considered so far captures many properties of the initially cold virial oscillations. However, the oscillations predicted by Eqs.\ \eqref{eq:xy_LO_ham} would never damp: damping is not included in this description\footnote{This feature is shared by other effective descriptions of virial oscillations like the so-called ``envelope equations'' derived by Levin et al.\ \cite{LevinEtAlphysrep:2014} to model the virial oscillations after water-bag initial conditions.}. This notwithstanding, one may assume that the damping, due to the back-action of the higher-order moments neglected so far, can be modeled by just adding the simplest dissipative term to Eqs.\ \eqref{eq:xy_LO_ham}, that is, by adding a term proportional to $-\dot x$ to Eq.\ \eqref{eq:ddotx}. Although, as we shall see in Sec.\ \ref{subsec_dissipation}, such an effective dissipation should be accompanied by a renormalization of the effective potential, this crude approximation indeed yields reasonable results for the asymptotic magnetization after the damping of the oscillations, that would simply coincide with the value $\overline{m}$ of $m$ corresponding to the minimum of $V_{\text{eff}}$. The first derivative of $V_{\text{eff}}(x)$ has a root in
\beq
\overline{x} = \frac{2}{3}\left(2 - \sqrt{1 + 3m_0^2} \right)~,
\label{eq:xbar_LO}
\eeq
so that, using $x = 2\left( 1 - m\right)$, we get
\beq
\overline{m} = \frac{1+\sqrt{1 + 3m_0^2}}{3}~.
\label{eq:mbar_LO}
\eeq
This result implies that a cold collapse will always lead to a state with magnetization
larger than a finite value that is predicted as $\overline{m}_{\text{min}} = \frac{2}{3}$; the latter lower limit corresponds to the initial condition $m_0 = 0$. Moreover, $\overline{m}$ increases with $m_0$, eventually reaching the  already discussed fixed point $\overline{m} = m_0 = 1$. All these properties are observed in the numerical simulations. 
The estimate for the asymptotic magnetization given by Eq.\ \eqref{eq:mbar_LO}, that contains no adjustable parameters, is compared with a more refined prediction to be derived in Sec.\ \ref{subsec_dissipation} and with the results obtained in numerical simulations in Fig.\ \ref{fig:m}. It is apparent that Eq.\ \eqref{eq:mbar_LO} systematically overestimates all the numerical results, including $\overline{m}_{\text{min}}$, but the trend is the correct one.

As a further consistency check of what we have derived so far we note that the leading-order prediction \eqref{eq:mbar_LO} for the magnetization after the damping of the oscillations is not only consistent with the virial theorem, as it should be for any stationary state, but it could have been directly derived from the virial theorem itself, exploiting the fact that the leading order corresponds to a harmonic approximation for the interparticle potential (see appendix \ref{app:virial} for details). 

\subsection{Effective dissipation}
\label{subsec_dissipation}
Let us now discuss how to effectively tackle the contribution of all the higher-order moments we have neglected so far and to show that such a contribution indeed yields an effective damping as well as a renormalization of the effective potential \eqref{eq:veff_cold}. 

We start by observing that, were the magnetization a given constant $m = \mu$ instead of a self-consistent function of time, and in the harmonic approximation that amounts to set $J = 0$ in Eqs.\ \eqref{eq:Idot}, the equations of motion for the moments would become a set of independent linear systems of differential equations; each system would be composed of $2L + 1$ coupled equations, where $2L = k+n$ is a non-negative even number, given by
\beq
\dot I_{k,n} = k \, I_{k-1,n+1} - n \mu \, I_{k+1,n-1}\,.
\label{eq:Idot_harm_mu}
\eeq
The real part of the eigenvalues of the matrix\footnote{Being the interaction among the $I$'s in Eqs.\ \eqref{eq:Idot_harm_mu} only between nearest neighbours, these matrices are tridiagonal and thus diagonalizable.} associated to each system vanishes. In fact Eqs.\ \eqref{eq:Idot_harm_mu} stem from the Vlasov equation for the HMF model \eqref{eq:Vlasov_HMF_symm} in the harmonic approximation (i.e., with $\sin\vartheta \approx \vartheta$) and with $m = \mu$. In this case the evolution in phase space reduces to a rigid rotation, i.e., to a periodic motion. Also the dynamics of the $I$'s must then be periodic and this rules out the possibility of a non-imaginary eigenvalue. Moreover, being the equations real, the conjugate of any eigenvalue is still an eigenvalue and since their number is odd there must be a zero mode. This means that the corresponding left eigenvector does not evolve and we have a first integral, linear in the $I_{k,n}$'s. Summarizing, we have a $2L$-dimensional effective linear dynamics whose eigenvalues are couples of imaginary conjugate numbers: we can thus consider any of the systems of $2L + 1$ equations with $k + n = L$ in Eqs.\ \eqref{eq:Idot_harm_mu} as the Hamilton equations of $2L$ decoupled harmonic oscillators. In addition, the periodicity implies that all the frequencies of the oscillators are integer multiples of a common fundamental one.
When we take into account the dependence of $m$ on the moments $I_{k,0}$ the equations become nonlinear and this simple picture is no longer valid. Nonetheless, being the equations invariant under time reversal, it is reasonable to assume the $L > 1$ moments to act as coupled mechanical systems exhibiting an oscillatory behaviour in the proximity of equilibrium. However, since the periodicity is lost, no assumption can be made on the values of the actual frequencies. 

In close analogy with the works by Caldeira and Leggett \cite{CaldeiraLeggett:prl1981,CaldeiraLeggett:physa1983} aiming at modeling a quantum dissipative environment in order to describe quantum brownian motion, we assume that the action of the high-order moments can be considered as that of a bath of independent harmonic oscillators coupled to the low-order ones. In this classical counterpart of the Caldeira-Leggett mechanism, dissipation emerges as the result of the loss of coherence of the set of oscillators, whose dynamics is individually reversible, when the number of oscillators diverges. The effective dissipation mechanism is thus seen as a ``classical decoherence''. It is worth noting that this feature is not exclusive of the Caldeira-Leggett model: at the classical level also other approaches exist where an effective dissipation may emerge from the coupling with an infinite assembly of oscillators (see e.g.\ \cite{Zwanzig:book}).  

Let us now go into the details. We retain the same notation as before, i.e., $x =  I_{0,2}$, $y = I_{1,1}$ and $z = I_{0,2}$, and we set 
\beq
m = 1 - \frac{x}{2} + \xi\,,
\eeq
where $\xi$ sums up all the contributions of the higher-order moments and is implicitly assumed small. The equations of motion \eqref{eq:xyz_LO} for $x$, $y$ and $z$ would then become
\begin{subequations}
\begin{align}
\dot x & =  2y\,, \label{eq:dotx_xi}\\
\dot y & =  z - \left(1 - \frac{x}{2} + \xi\right) x\,, \label{eq:doty_xi}\\
\dot z & =  -2 \left(1 - \frac{x}{2} + \xi\right) y\, \label{eq:dotz_xi},
\end{align}
\label{eq:xyz_LO_xi}
\end{subequations}
but being $\xi$ an (unknown) function of time we cannot obtain an integral of motion by manipulating these equations as above. Still, we impose the conservation of energy 
\beq
\varepsilon = \frac{z}{2} - \frac{1}{2}\left( 1 - \frac{x}{2} + \xi\right)^2
\eeq
so that we can express $z$ as a function of $\varepsilon$, $x$ and $\xi$ and eliminate it from the equations \eqref{eq:xyz_LO_xi}, that become
\begin{subequations}
\begin{align}
\dot x & =  2y\,, \label{eq:dotx_ham_xi}\\
\dot y & =  2\varepsilon + \left( 1 - \frac{x}{2} + \xi\right)^2 - \left( 1 - \frac{x}{2} + \xi\right) x \,. \label{eq:doty_ham_xi}
\end{align}
\label{eq:xy_LO_ham_xi}
\end{subequations}  
Introducing 
\beq
g(x)=x(2-x)\,, 
\eeq
such that 
\beq
\frac{dg}{dx} = g'(x) = 2(1-x)\,, 
\eeq
and neglecting $\mathcal{O}(\xi^2)$ terms, Eqs.\ \eqref{eq:xy_LO_ham_xi} are equivalent to 
\beq
\frac{1}{2}\ddot x = -\frac{d}{dx} V_{\text{eff}}(x) +\xi\, g'(x)\,,
\label{eq:ddotx_xi}
\eeq
where $V_{\text{eff}}(x)$ is still given by Eq.\ \eqref{eq:veff}. We have now to describe the dynamics of $\xi$. We assume that $\xi$ is a sum of many oscillating terms,
\beq
\xi = -\zeta_0  g(x) + \sum_{k = 1}^M c_k q_k\,, 
\label{xi_CL}
\eeq
where the $q_k$'s are the coordinates of a set of $M$ harmonic oscillators, with $M$ large; $\zeta_0$ and $c_1,\ldots,c_M$ are real constants. 
The first term in Eq.\ \eqref{xi_CL} implies that the coupling between low- and high-order modes may modulate the average values of the oscillations. As we shall see in the following, we can think of such a term as a counter-term that tunes the renormalization of the effective potential, allowing to make a simplifying assumption on the spectrum of the oscillators. 
Equation \eqref{eq:ddotx_xi} then becomes
\beq
\frac{1}{2}\ddot x = -\frac{d}{dx} V_{\text{eff}}(x) - \zeta_0 g'(x)g(x) + g'(x) \sum_{k = 1}^M c_k q_k\,.
\label{eq:ddotx_xi_bis}
\eeq
To derive the equations of motion for the oscillators, we assume the above equation derives from a Lagrangian $L$ given by
\beq
L = L_x + L_{\text{bath}} + L_{\text{int}} + L_{\text{ct}}\, 
\label{eq:L}
\eeq 
where
\beq
L_x = \frac{1}{4}\dot x^2 - V_{\text{eff}}(x)
\eeq
describes the $x$ sector, i.e., the dynamics of the lowest-order modes,
\beq
L_{\text{bath}} = \frac{1}{2}\sum_{k = 1}^M \left( \dot q_k^2 - \omega_k^2 q_k^2 \right) 
\eeq
describes the bath of harmonic oscillators modeling the higher-order moments,
\beq
L_{\text{int}} = g(x)\sum_{k = 1}^M c_k q_k
\eeq
describes the interaction between the bath and the lowest-order moments\footnote{Such an interaction is linear in the $q_k$'s as in the classical Caldeira-Leggett model, but at variance with the latter model the coupling with the $x$ degree of freedom is nonlinear and realized via the function $g(x)$.}, and
\beq
L_{\text{ct}} = -\frac{1}{2}\zeta_0 g^2(x)
\eeq
corresponds to the counter-term. From the Lagrangian \eqref{eq:L} not only Eq.\ \eqref{eq:ddotx_xi_bis} follows, but also the equations of motion of the bath degrees of freedom, as
\beq
\ddot q_k = - \omega_k^2 q_k + c_k g(x)\,,
\label{eq:ddotq}
\eeq
with $k = 1,\ldots,M$. From Eq.\ \eqref{eq:ddotq} we see that the low-order modes act on the oscillators of the bath by displacing their equilibrium position. Let us now solve the dynamics of the oscillators of the bath: this can be done thanks to the linearity in $q_k$ of Eqs.\ \eqref{eq:ddotq}. As shown in Appendix \ref{app:calculations}, we get, for $k = 1,\ldots,M$, 
\beq
q_k(t) = q_k^0(t) + \frac{c_k}{\omega_k^2} g\left(x(t)\right) - \frac{c_k}{\omega_k^2} \frac{d}{dt} \int_0^t g\left(x(\tau) \right) \cos\left[\omega_k (t - \tau) \right]\, d\tau\,,
\label{eq:qk}
\eeq
where the first term on the r.h.s.\ is the homogeneous solution
\beq
q^0_k(t) = \frac{\dot q_k(0)}{\omega_k} \cos(\omega_k t) + q_k(0) \sin(\omega_k t)\,,
\label{eq:q0}
\eeq
while the other two terms come from the interaction. Using Eq.\ \eqref{eq:qk} we can write
\beq
\sum_{k = 1}^M c_k q_k = \sum_{k = 1}^M c_k q^0_k(t) + g(x) \sum_{k = 1}^M \frac{c^2_k}{\omega_k^2} -  \frac{d}{dt} \int_0^t g\left(x(\tau) \right) \sum_{k = 1}^M \frac{c^2_k}{\omega_k^2}\cos\left[\omega_k (t - \tau) \right]\, d\tau\,.
\label{eq:sumckqk}
\eeq
We now want to take the limit of a large number of oscillators, $M\to\infty$. In order to replace the sums with integrals over the frequencies we introduce the spectral density $J(\omega)$ defined as
\beq
J(\omega) = \frac{\pi}{2}\sum_{k = 1}^\infty \frac{c^2_k}{\omega_k} \delta(\omega - \omega_k)  
\eeq
so that Eq.\ \eqref{eq:sumckqk} becomes
\beq
\sum_{k = 1}^\infty c_k q_k = \xi^0(t) + \frac{2}{\pi}g(x) \int_0^\infty d\omega \,\frac{J(\omega)}{\omega} - \frac{2}{\pi} \frac{d}{dt} \int_0^t d\tau \, g\left(x(\tau) \right) \int_0^\infty d\omega \, \frac{J(\omega)}{\omega}\cos\left[\omega (t - \tau) \right]\,,
\label{eq:intckqk}
\eeq
where 
\beq
\xi^0(t) = \sum_{k = 1}^\infty c_k q^0_k(t) \,.
\eeq
To go further we need some assumptions on the spectral density. As in the standard Caldeira-Leggett model we assume a white spectrum up to a cutoff $\Omega$, that is, an Ohmic spectral function,
\beq
\frac{J(\omega)}{\omega} = \left\{\begin{array}{ccl} \eta & & 
\omega \le  \Omega\,; \\
0  & & 
\omega > \Omega\,,
\end{array} 
\right.
\eeq
and substituting the latter into Eq.\ \eqref{eq:intckqk} we get
\beq
\sum_{k = 1}^\infty c_k q_k = \xi^0(t) + \frac{2\eta \Omega}{\pi}g(x) - \eta\, \frac{d}{dt} \int_0^t d\tau \, g\left(x(\tau) \right) \frac{2}{\pi} \int_0^\Omega d\omega \, \cos\left[\omega (t - \tau) \right] \,.
\label{eq:intcketa}
\eeq
The second term in the r.h.s.\ of the above equation, that is, the displacement of the oscillators due to the coupling, is proportional to $\Omega$, so that it would diverge in the $\Omega \to \infty$ limit. However, the quantity given by Eq.\ \eqref{eq:intcketa} is not an observable quantity, at variance with $\xi$; inserting Eq.\ \eqref{eq:intcketa} into Eq.\ \eqref{xi_CL} we get 
\beq
\xi = \xi^0(t) + \left(\frac{2\eta}{\pi} \Omega - \zeta_0\right) g(x) - \eta\, \frac{d}{dt} \int_0^t d\tau \, g\left(x(\tau) \right) \frac{2}{\pi} \int_0^\Omega d\omega \, \cos\left[\omega (t - \tau) \right] \,,
\label{eq:pippo}
\eeq 
so that we can reabsorb the divergence into the bare coupling constant $\zeta_0$, and by taking the limit $\Omega\to\infty$ we get
\beq
\xi = \xi^0(t) - \zeta g(x) - \eta\, \frac{d}{dt} \int_0^t d\tau \, g\left(x(\tau) \right) \delta(t - \tau)\,,
\label{eq:xinew}
\eeq
where $\zeta$ is the renormalized (and thus unknown) coupling
\beq
\zeta = \lim_{\Omega\to\infty} \left(\zeta_0 - \frac{2\eta}{\pi} \Omega\right) 
\eeq
and where we have used the integral representation of the Dirac $\delta$ distribution. By interchanging the derivative and the integral in Eq.\ \eqref{eq:xinew} we obtain
\beq
\xi = \xi^0(t) - \zeta g(x) - \eta g'(x) \,\dot x\, ,
\label{eq:xirenorm}
\eeq
so that, after substituting Eq.\  \eqref{eq:xirenorm} into \eqref{eq:ddotx_xi}, the equation of motion for $x$ finally reads
\beq
\frac{1}{2}\ddot x = -\frac{d}{dx} V_{\text{eff}}(x) - \zeta g(x)g'(x) - \eta \left[g'(x)\right]^2 \,\dot x + g'(x)\,\xi^0(t)\, .
\label{eqmotox_xi}
\eeq
Being $\eta \left[g'(x)\right]^2 \ge 0$, the third term in the r.h.s.\ of Eq.\ \eqref{eqmotox_xi} is indeed a damping term, although depending also on $x$. Given that $\xi_0(t)$ is an infinite sum of oscillating terms with zero mean and a white spectrum, each term being given by Eq.\ \eqref{eq:q0}, it is equivalent to a white noise with zero mean. Hence Eq.\ \eqref{eqmotox_xi} is a Langevin equation with multiplicative noise, at variance with the classical Caldeira-Leggett model where the noise is purely additive. The r.m.s.\ amplitude of $\xi^0(t)$ depends on the initial conditions of the oscillators; were the latter in thermal equilibrium at a given temperature $T$, as in the classical Caldeira-Leggett model, one would have $\langle \xi^0(t)\xi^0(t') \rangle \propto \eta T \delta(t - t')$, in agreement with the fluctuation-dissipation theorem \cite{CaldeiraLeggett:prl1981,CaldeiraLeggett:physa1983}. However, in our case there is no thermal bath, and the oscillators only model the contribution of the higher-order moments, that is assumed to be small for our theory to be consistent.
We thus make the further assumption that the initial conditions of the oscillators are such that the amplitude of the noise (or the effective temperature of the bath) is very small, so that we can neglect the multiplicative noise in the following. 

\begin{figure}
\includegraphics[width = 0.7\textwidth]{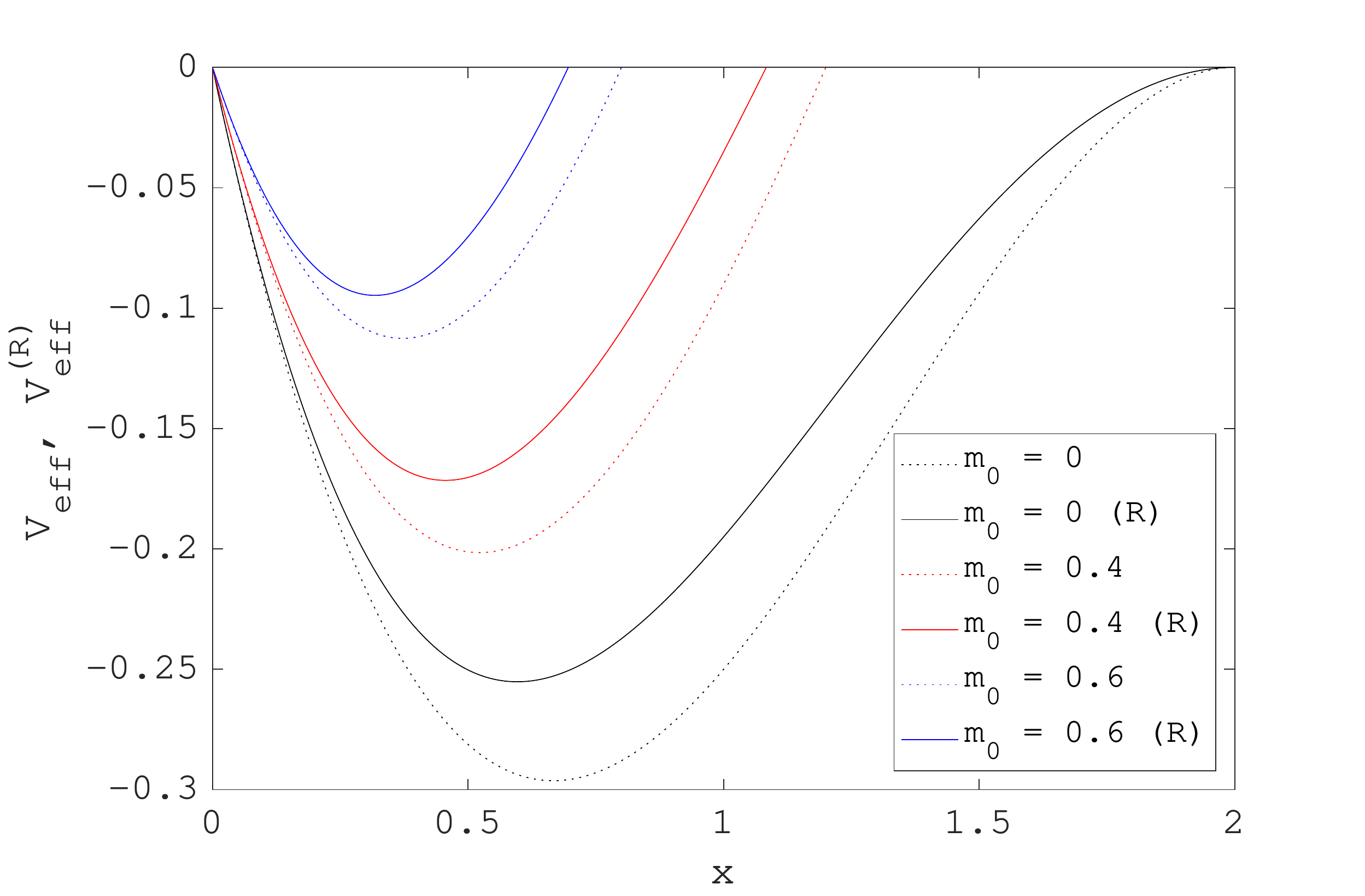}
\caption{Comparison between the bare effective potential $V_{\text{eff}}$ given by Eq.\ \eqref{eq:veff_cold} (dotted lines) and the renormalized potential $V^{(R)}_{\text{eff}}$ given by Eq.\ \eqref{eq:Veff_R} (solid lines) as a function of $x$ in the case of cold collapse for some values of the initial magnetization $m_0$ (see inset). Here $\zeta = 0.11$.}
\label{fig:veff_R}  
\end{figure}

\begin{figure}
\includegraphics[width = 0.7\textwidth]{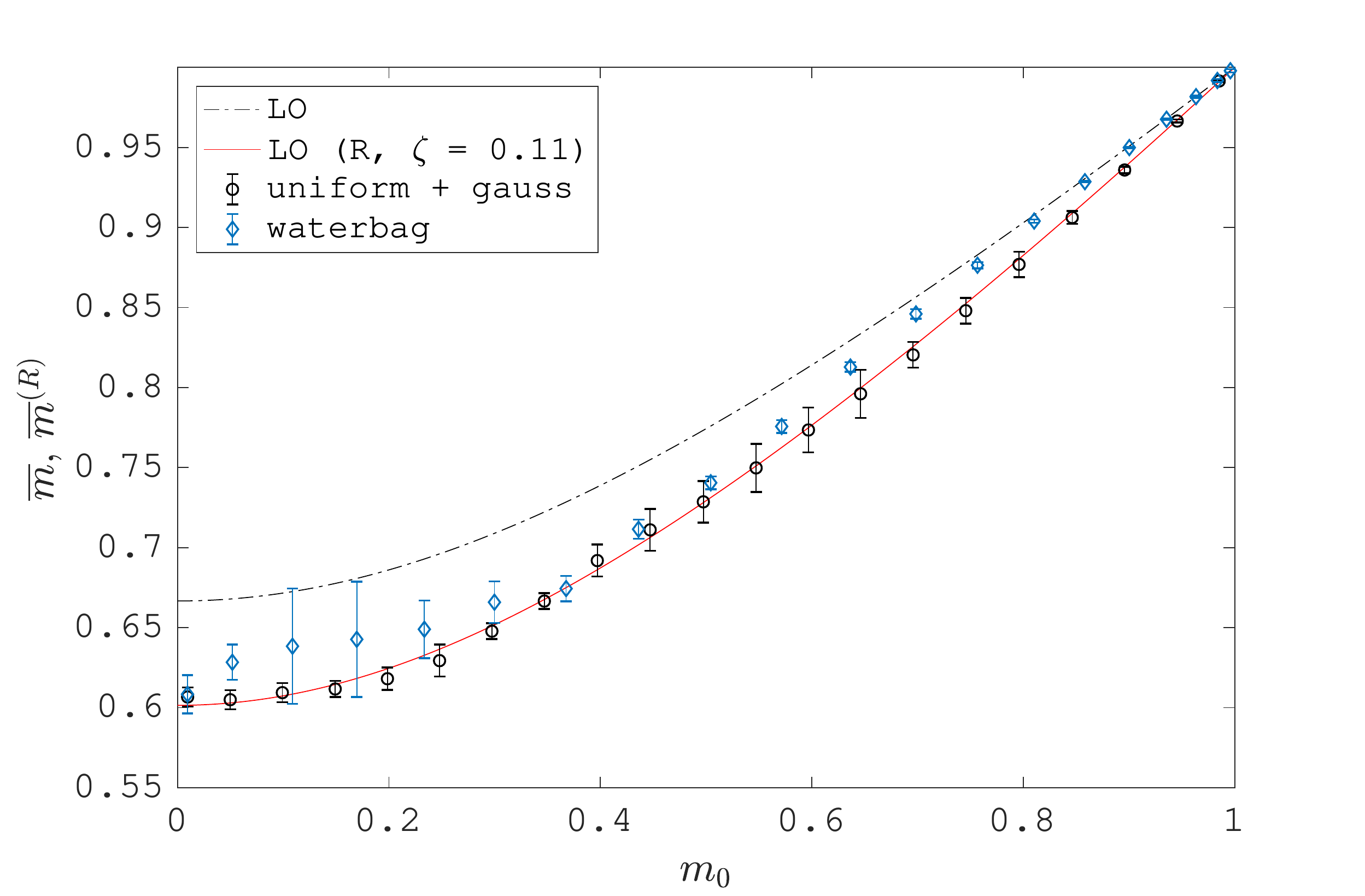}
\caption{Comparison between the leading-order theoretical predictions for the asymptotic magnetization after a cold collapse and the numerical results. The black dash-dotted curve is the bare $\overline{m}$ given by Eq.\ \eqref{eq:mbar_LO} that does not depend on any adjustable parameters; the red solid curve is the leading-order prediction with the dissipative correction $\overline{m}^{(R)}$ given by Eqs.\ \eqref{eq:mplusdeltam} and \eqref{eq:deltam_LO_R}, with $\zeta = 0.11$; symbols with errorbars refer to results obtained with numerical simulations of the cold collapse (vanishing initial kinetic energy) of an HMF model with $N = 2\times 10^5$ particles (black open circles correspond to initial conditions where a Gaussian overdensity is superimposed to a uniform spatial distribution in $[-\pi,\pi]$; blue open diamonds correspond to ``waterbag'' initial conditions, where the particles are uniformly distributed in $[-a,a]$ with $|a| \le \pi$).}
\label{fig:m}  
\end{figure}

The multiplicative nature of the noise and the dependence on $x$ of the damping term are not the only differences between Eq.\  \eqref{eqmotox_xi} and the Langevin equation of the classical Caldeira-Leggett model: there is also the term involving the coupling $\zeta$, that can be seen as a ``dissipative correction'' that renormalizes the effective potential $V_{\text{eff}}$. After neglecting the noise term and recalling that $g(x) = x(2-x)$ we can indeed write Eq.\ \eqref{eqmotox_xi} as
\beq
\frac{1}{2}\ddot x = -\frac{d}{dx} V^{(R)}_{\text{eff}}(x) - 4 \eta \left(1-x\right)^2 \dot x\,,
\label{eqmotox_xi_final}
\eeq
where
\beq
V^{(R)}_{\text{eff}}(x) = V_{\text{eff}}(x) + \frac{\zeta}{2}x^2 (2 - x)^2
\label{eq:Veff_R}
\eeq
is the renormalized effective potential. As observed above, the value of the parameter $\zeta$, as well as that of the friction coefficient $\eta$, can not be predicted by the theory. Yet, they both have to be small for the theory to be consistent. Moreover, Eq.\ \eqref{eq:Veff_R} implies that the dissipative correction to the effective potential vanishes when $x=0$ and $x=2$, regardless of the value of $\zeta$, and this is physically sound, because these values of $x$ correspond to $m = 1$ and $m=0$, respectively, and are fixed points of the dynamics of the system, that are thus left unchanged by the dissipative correction. A comparison between the bare and renormalized effective potentials is shown in Fig.\ \ref{fig:veff_R}.
Equation \eqref{eqmotox_xi_final} implies that the asymptotic values of $x$, that determine those of the magnetization $m$, are the minima $\overline{x}^{(R)}$ of $V^{(R)}_{\text{eff}}(x)$. To calculate them we exploit the fact that the dissipative correction is small, writing $\overline{x}^{(R)} = \overline{x} + \delta \overline{x}$ where 
$\delta \overline{x} = \mathcal{O}(\zeta)$ and expanding $V'^{(R)}_{\text{eff}}(x)$ around $\overline{x}$ up to first order in $\zeta$, obtaining
\beq
\delta \overline{x} = -\zeta\,\frac{g(\overline{x}) g'(\overline{x})}{2 - \frac{3}{2} \overline{x}}~.
\label{eq:deltax}
\eeq
Equation \eqref{eq:deltax} implies that the dissipative correction $\delta \overline{m}$ to the leading-order estimate of the magnetization $\overline{m} = 1 -\overline{x}/2$ is 
\beq
\delta \overline{m} = -\zeta\, \overline{x} \frac{\left(2 - \overline{x} \right)^2}{4 - 3 \overline{x}}~,
\eeq
so that if $\zeta > 0$ we have a negative correction, that is the correct one since, as already observed, $\overline{m}$ overestimates the numerical results. Using the expression of $\overline{x}$ given by Eq.\ \eqref{eq:xbar_LO}, the the leading-order prediction of the asymptotic magnetization $\overline{m}^{(R)}$, including the dissipative correction, as a function of the initial magnetization $m_0$ is thus
\beq
\overline{m}^{(R)} = \overline{m} + \delta \overline{m}
\label{eq:mplusdeltam}
\eeq
where $\overline{m}$ is given by Eq.\ \eqref{eq:mbar_LO} and  
\beq
\delta \overline{m} = -\frac{4}{27} \zeta \, \frac{(2 - \Delta)(1 + \Delta)^2}{\Delta}
\label{eq:deltam_LO_R}
\eeq 
with
\beq
\Delta = \sqrt{1 + 3 m_0^2}~. 
\label{eq:Delta}
\eeq
The leading-order prediction of the asymptotic magnetization $\overline{m}$, given by Eqs.\ \eqref{eq:mplusdeltam} and \eqref{eq:deltam_LO_R} is plotted in Fig.\ \ref{fig:m} against the initial magnetization $m_0$ and compared with the leading-order prediction without the dissipative correction given by Eq.\ \eqref{eq:mbar_LO} and the results of numerical simulations of cold collapse of an HMF model with $N = 2\times 10^5$ particles, for two different classes of initial conditions. Choosing $\zeta = 0.11$ the agreement between the theoretical prediction and the simulated data is very good, although the leading-order prediction depends only on $m_0$ so that it cannot resolve the fine differences in the asymptotic magnetization obtained with different initial conditions corresponding to the same $m_0$.  

\begin{figure}
\includegraphics[width = 0.7\textwidth]{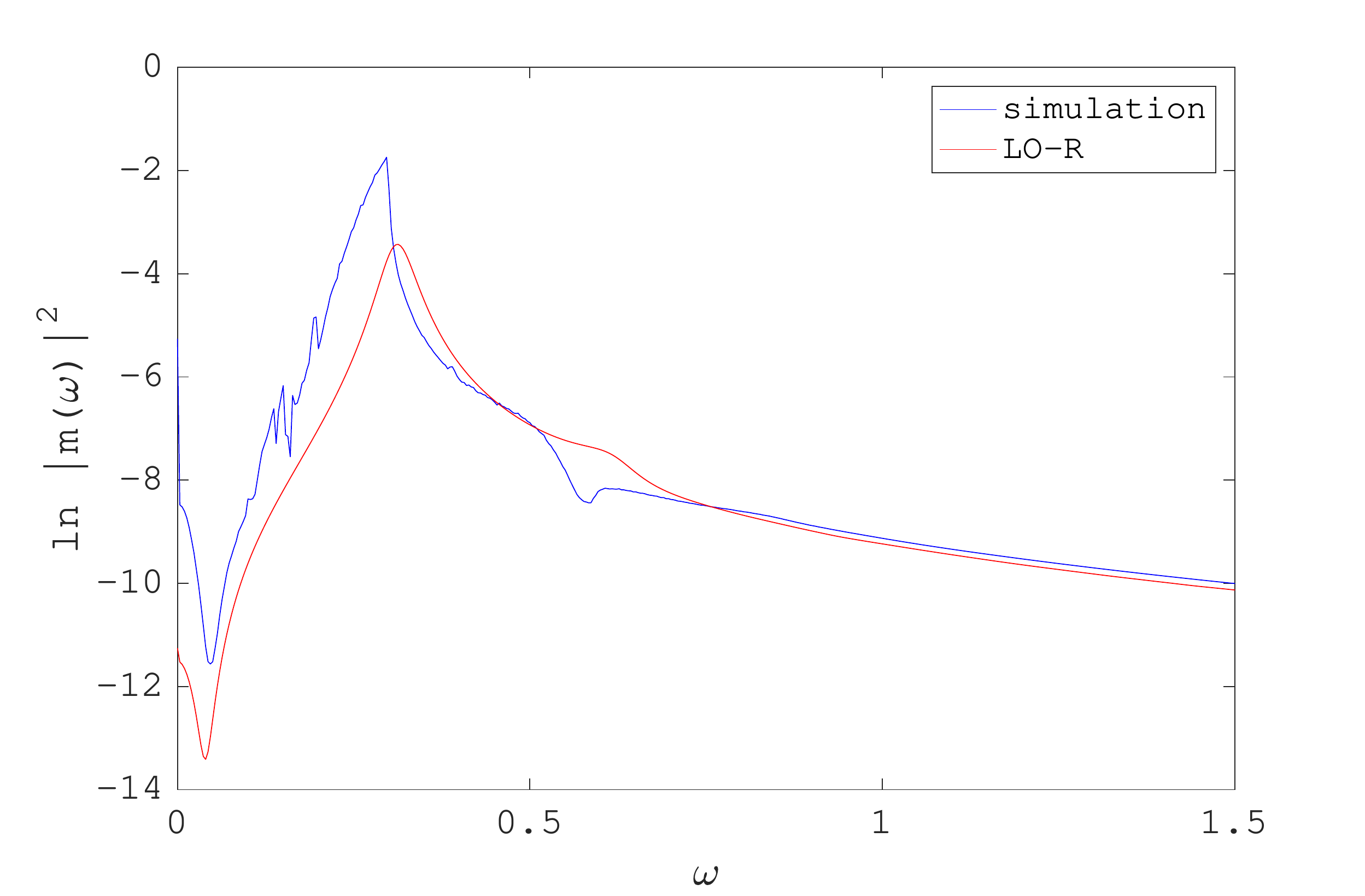}
\caption{Comparison between the Fourier power spectra of the solution of Eq.\ \eqref{eqmotox_xi_final}, with $\zeta = 0.11$ and $\eta = 0.1$ (red line), and of the $m(t)$ obtained with a numerical simulation of an HMF model with $N = 2\times 10^6$ particles (blue line). Here $m_0 = 0.775$ and the initial conditions of the simulation are uniform with a Gaussian overdensity. The vertical scale is logarithmic.}
\label{fig:fourier_LO}  
\end{figure}

Going beyond the prediction of the asymptotic value of the magnetization, we do not expect the solutions of Eq.\ \eqref{eqmotox_xi_final} to be able to accurately reconstruct the actual time evolution $m(t)$ during the cold collapse; although Eq.\ \eqref{eqmotox_xi_final} contains the damping, it still has only the information on the lowest-order moments. Indeed, when $m_0$ is small the agreement between theoretical and numerical outcomes of $m(t)$ is only qualitative, but when $m_0$ is large enough, so that the system is almost always in a tightly collapsed state, the Fourier power spectra of the solutions of Eq.\ \eqref{eqmotox_xi_final} are not very different from the ones of the $m(t)$ obtained in numerical simulations: an example is shown in Fig.\ \ref{fig:fourier_LO}. This is a further confirmation of the fact that our effective description captures a non-negligible part of the actual dynamics already at the leading order.
 
It is worth mentioning that the relation between dissipation in Vlasov dynamics and the Caldeira-Leggett model had been previously discussed by Hagstrom and Morrison \cite{HagstromMorrison:physicad2011} but only in the special context of linear Landau damping in a homogeneous background. More precisely, they found an explicit mapping between the Vlasov-Poisson system and the Hamilton equations of the Caldeira-Leggett model, by expressing both systems in terms of normal modes. Although less rigorous than Hagstrom and Morrison's, our results suggest that the analogy between collisionless dissipation and the Caldeira-Leggett mechanism can be pushed considerably forward to include the fully nonlinear case of violent relaxation towards non-homogeneous states.  

\subsection{Next-to-leading order description of cold collapse}
\label{subsec_next}

The time evolution of the inertia moments of the distribution function can be studied, in principle, up to any desired order. Going to the next-to-leading order in the dynamics of the $I_{k,n}$ means considering $J = 1$ and $k+n \le 4$. This involves eight moments, including the three already appearing at the leading order, and the equations of motion of these moments, neglecting all the higher-order ones, read as
\begin{subequations}
\begin{align}
\dot I_{2,0} & = 2I_{1,1}  \\
\dot I_{1,1} & = I_{0,2} - m\left(I_{2,0} - \frac{1}{6} I_{4,0}\right) \\
\dot I_{0,2} & = - 2m\left(I_{1,1} - \frac{1}{6} I_{3,1}\right) \\
\dot I_{4,0} & = 4I_{3,1}  \\
\dot I_{3,1} & = 3I_{2,2} - mI_{4,0}  \\
\dot I_{2,2} & = 2I_{1,3} - 2mI_{3,1}  \\
\dot I_{1,3} & = I_{0,4} - 3mI_{2,2}  \\
\dot I_{0,4} & = - 4mI_{1,3} 
\end{align}
\label{eq:Idot_NLO}
\end{subequations}
where
\beq
m = 1 - \frac{1}{2}I_{2,0} + \frac{1}{24}I_{4,0}~. 
\label{eq:m_NLO}
\eeq  
Equations \eqref{eq:Idot_NLO} are invariant under time reversal, so that they cannot exhibit any damping and relaxation towards a fixed point. However, as in the case of the leading order, we expect the higher-order terms here neglected to provide an effective dissipation that we can model, in the simplest way, by adding dissipative terms to Eqs.\ \eqref{eq:Idot_NLO} as follows:
\begin{subequations}
\begin{align}
\dot I_{2,0} & = 2I_{1,1}  \\
\dot I_{1,1} & = I_{0,2} - m\left(I_{2,0} - \frac{1}{6} I_{4,0}\right) -\gamma_{(2)}  I_{1,1}\\
\dot I_{0,2} & = - 2m\left(I_{1,1} - \frac{1}{6} I_{3,1}\right) \\
\dot I_{4,0} & = 4I_{3,1}  \\
\dot I_{3,1} & = 3I_{2,2} - mI_{4,0}  -\gamma_{(4)}  I_{3,1}\\
\dot I_{2,2} & = 2I_{1,3} - 2mI_{3,1}  \\
\dot I_{1,3} & = I_{0,4} - 3mI_{2,2}  -\gamma_{(4)}  I_{1,3}\\
\dot I_{0,4} & = - 4mI_{1,3} 
\end{align}
\label{eq:Idot_NLO_diss}
\end{subequations}
where $m$ is still given by Eq.\ \eqref{eq:m_NLO} and the real positive constants $\gamma_{(L)}$, with $L = 2$ and $L = 4$, respectively, are the effective friction coefficients that are expected to be different for moments of different order: in particular, since the dissipation time scale of the higher-order moments should be longer than that of the lower-order moments, $\gamma_{(2)} > \gamma_{(4)}$. In Eqs.\ \eqref{eq:Idot_NLO_diss} we could have considered also off-diagonal dissipative terms, but since this is only an effective description and we do not expect the precise form of the dissipative terms to dramatically affect the main features of the dynamics, we chose the simplest possible form. We note that even with the inclusion of the friction terms the energy per particle $\varepsilon$ is conserved, because
\beq
\frac{d\varepsilon}{dt} = \frac{1}{2}\frac{d}{dt} \left(I_{0,2} - m^2\right) = \frac{1}{2}\dot I_{0,2} - m\left(-\frac{1}{2}\dot I_{2,0} +\frac{1}{24}\dot I_{4,0}\right)  = - m\left(I_{1,1} - \frac{1}{6} I_{3,1}\right) +m \left(I_{1,1} - \frac{1}{6} I_{3,1}\right) =0\,.
\eeq 
In principle we expect that a proper effective treatment of the higher-order moments would give rise not only to dissipative terms but also to a renormalization of the dynamics, like the dissipative correction to the effective potential arising at the leading order: however, since an analytical approach like that carried out in Sec.\ \ref{subsec_dissipation} appears extremely difficult if not unfeasible at the next-to-leading order, we neglected this aspect and numerically solved Eqs.\ \eqref{eq:Idot_NLO_diss} for different initial conditions, estimating the asymptotic values of $m$ from the numerical solution. Therefore we cannot expect the prediction of Eqs.\ \eqref{eq:Idot_NLO_diss} to accurately match numerical data: yet, if our approach is consistent, such a prediction should be closer to the numerical data than the bare leading-order prediction given by Eq.\ \eqref{eq:mbar_LO}. Moreover, while the asymptotic magnetization predicted at the leading order depends only on the initial magnetization $m_0$, here a dependence on finer details of the initial conditions may show up.

The space of the initial conditions of the system of equations \eqref{eq:Idot_NLO_diss} is eight-dimensional. However, considering only cold initial conditions implies that all the moments containing $p$ initially vanish, that is, the only moments that can be nonzero in the initial state are $I_{2,0}$ and $I_{4,0}$. The initial condition can thus be parametrized by two independent quantities. Observing that $\left\langle\left(\vartheta^2 - \langle\vartheta\rangle^2 \right)^2 \right\rangle \ge 0$ implies $I_{4,0} \ge I^2_{2,0}$ we can define a parameter $b \ge 1$ such as
\beq
I_{4,0} = b I^2_{2,0}
\eeq   
and parametrize the initial condition by means of $m_0$ and $b$. Since
\beq
m_0 = 1 - \frac{1}{2}I_{2,0} (0) + \frac{b}{24}I^2_{2,0}~,
\eeq 
given $m_0$ and $b$ the value of $I_{2,0}(0)$ is 
\beq
I_{2,0}(0) = \frac{6}{b} \left(1 - \sqrt{1 -\frac{2b}{3}(1 - m_0)} \right)~.
\label{eq:I0m0}
\eeq 
According to Eq.\ \eqref{eq:I0m0}, if $b<3/2$ all the values of $m_0$ are allowed; if $b > 3/2$ some values of $m_0$ are forbidden. For a uniform distribution $b = 1.8$, and for the initial conditions of the simulations considered in Fig.\ \ref{fig:m} (those obtained with a Gaussian overdensity superimposed on a uniform distribution) $b \in (1.8,9)$ as long as $m_0 \lesssim 0.8$ and becomes much larger for larger initial magnetizations. 

\begin{figure}
\includegraphics[width = 0.7\textwidth]{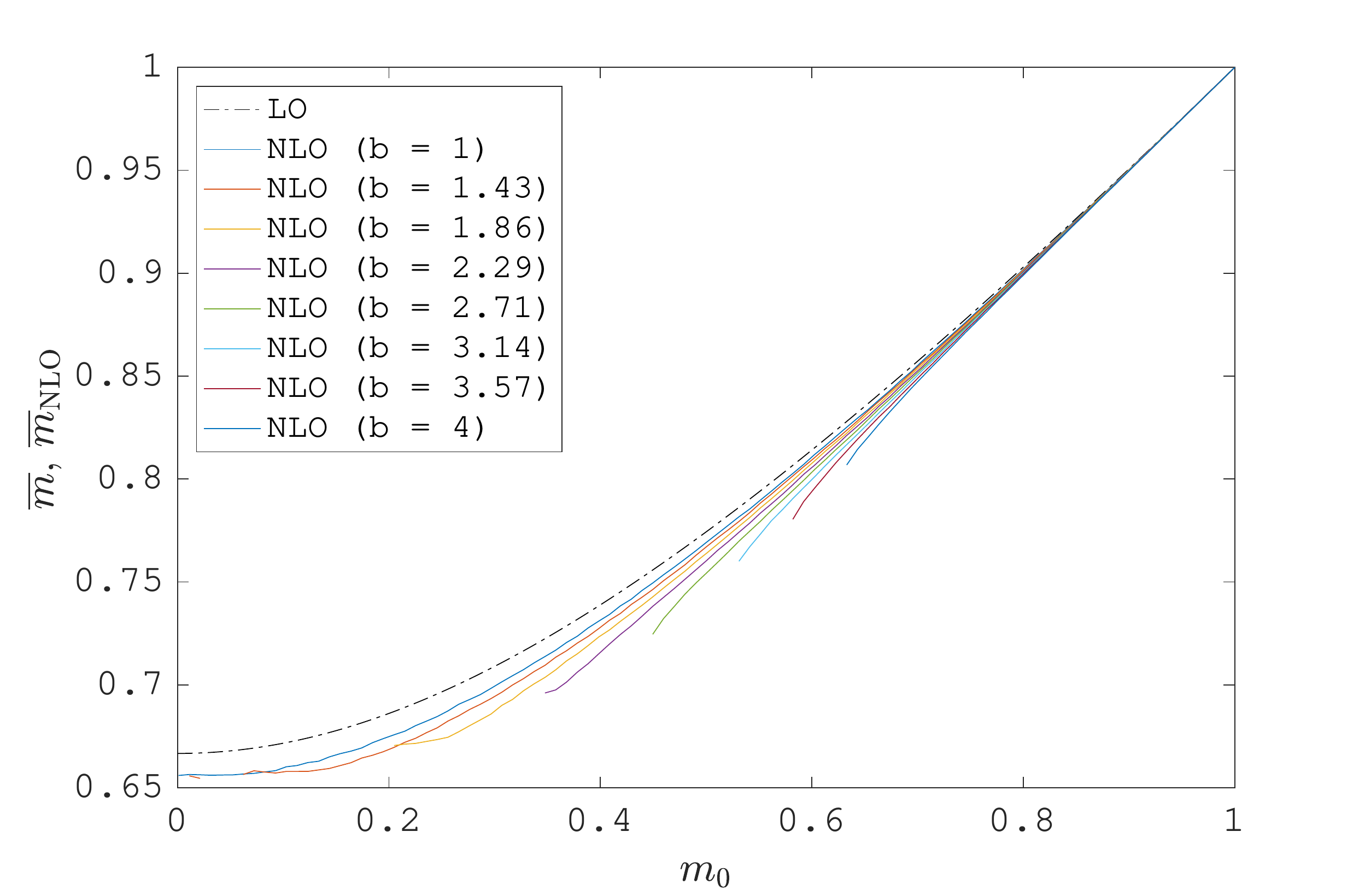}
\caption{Next-to-leading order results $\overline{m}_{\text{NLO}}$ for the asymptotic magnetization obtained by means of the numerical solution of Eqs.\ \eqref{eq:Idot_NLO_diss} with $\gamma_{(2)} = 0.1$ and $\gamma_{(4)} = 0.05$ (color curves for different values of $b$, see legend). The bare leading-order prediction $\overline{m}$ given by Eq.\ \eqref{eq:mbar_LO} (black dot-dashed curve) is also shown as a comparison.}
\label{fig:NLO}  
\end{figure}

\begin{figure}
\includegraphics[width = 0.7\textwidth]{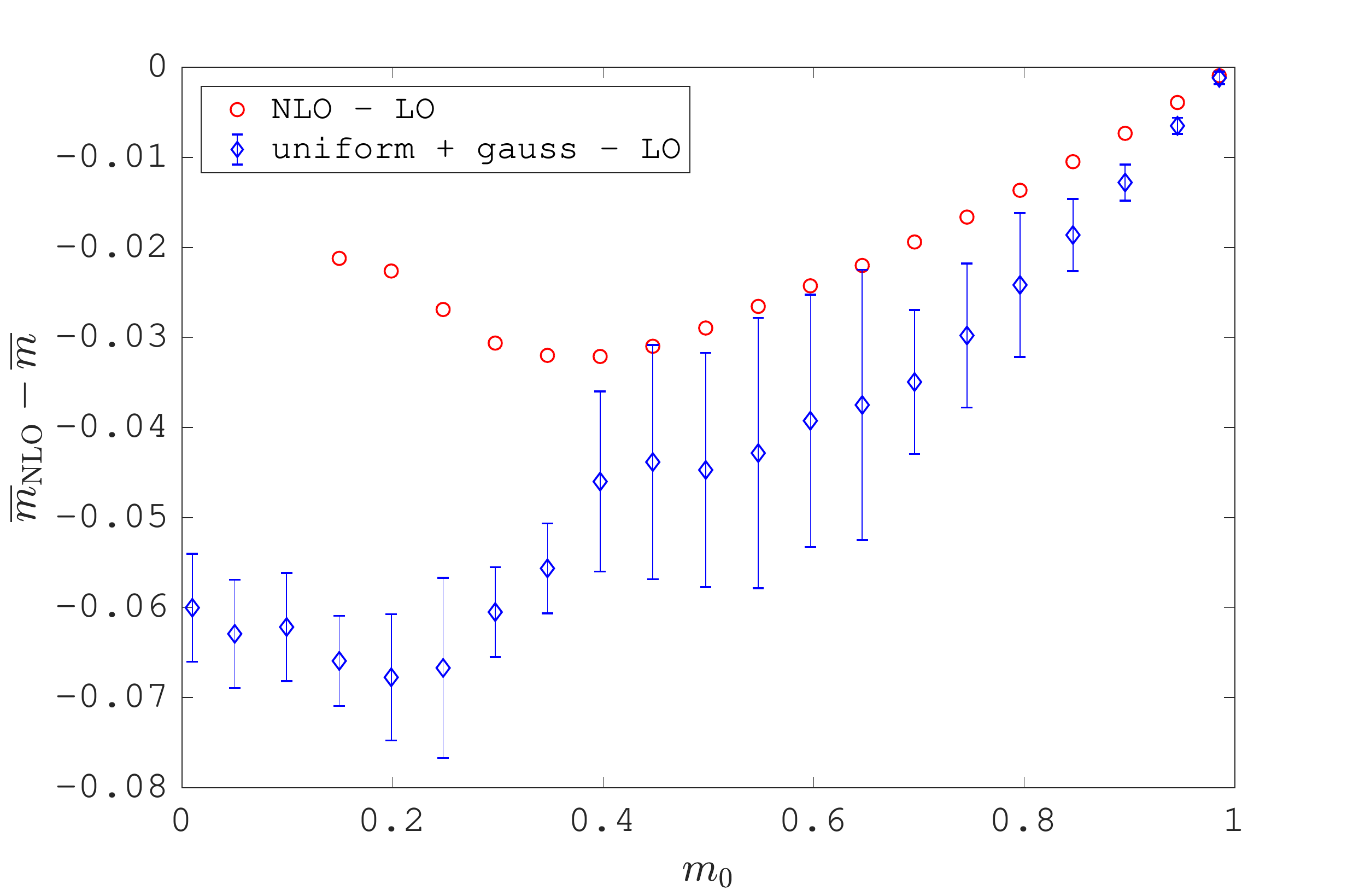}
\caption{Difference $\overline{m}_{\text{NLO}} - \overline{m}$ between the next-to-leading order and the bare leading-order prediction of the asymptotic magnetization (red circles) compared to the difference between simulation results obtained starting with uniform distributions plus a Gaussian overdensity and the leading order prediction (blue diamonds with errorbars; same data shown in Fig.\ \ref{fig:m}). The theoretical values have been computed solving Eqs.\ \eqref{eq:Idot_NLO_diss} with $\gamma_{(2)} = 0.1$ and $\gamma_{(4)} = 0.05$ and using the values of $b$ extracted from the initial conditions of the simulations at the corresponding initial magnetization $m_0$. Theoretical results for $m_0 \le 0.1$ are not shown because Eqs.\  \eqref{eq:Idot_NLO_diss} predict a vanishing asymptotic magnetization in these cases.}
\label{fig:delta_NLO}  
\end{figure}

Numerical results confirm that the next-to-leading order predictions of the asymptotic magnetization do not match the numerical data, yet systematically improve the bare leading-order $\overline{m}$ given by Eq.\  \eqref{eq:mbar_LO}. Moreover, a dependence on $b$ shows up, although rather weak: this is consistent with the fact that in a cold collapse the dominant effect is that of the lowest-order moments, so that the outcome is mostly determined by $m_0$. in Fig.\ \ref{fig:NLO} we plot the values of $\overline{m}_{\text{NLO}}$ obtained by means of the numerical solution of Eqs.\ \eqref{eq:Idot_NLO_diss} with $\gamma_{(2)} = 0.1$ and $\gamma_{(4)} = 0.05$ for some values of $b$, i.e., $b\in[1,4]$.  
In order to appreciate the quantitative improvement with respect to the bare leading-order result, in Fig.\ \ref{fig:delta_NLO} we show the difference $\overline{m}_{\text{NLO}} - \overline{m}$ between the next-to-leading order and the bare leading-order prediction for the appropriate values of $b$ relative to the initial conditions of a subset of the simulations results already shown in Fig.\ \ref{fig:m}, together with the difference between the outcome of the simulations and the bare leading order value. It is apparent that for $m_0 \gtrsim 0.4$ the next-to-leading order prediction substantially improves the base leading-order prediction, giving more than half of the needed correction to match the simulation data; for smaller initial magnetizations the improvement is less substantial.  
We note that for small values of $b$ and small values of $m_0$ we obtain a uniform asymptotic state, i.e., $m = 0$ (data not shown in Figs.\ \ref{fig:NLO} and \ref{fig:delta_NLO}). Given that when $\varepsilon < 0$ the magnetization can not vanish, this is a shortcoming of our theoretical approach\footnote{Or better of the numerical implementation we are considering now at the next-to-leding order.} that, as discussed above, is better suited for describing collapsed states, so that it may not work well for very small initial magnetizations. However, this concerns only a very small subset of the space of initial conditions. 

\begin{figure}
\includegraphics[width = 0.7\textwidth]{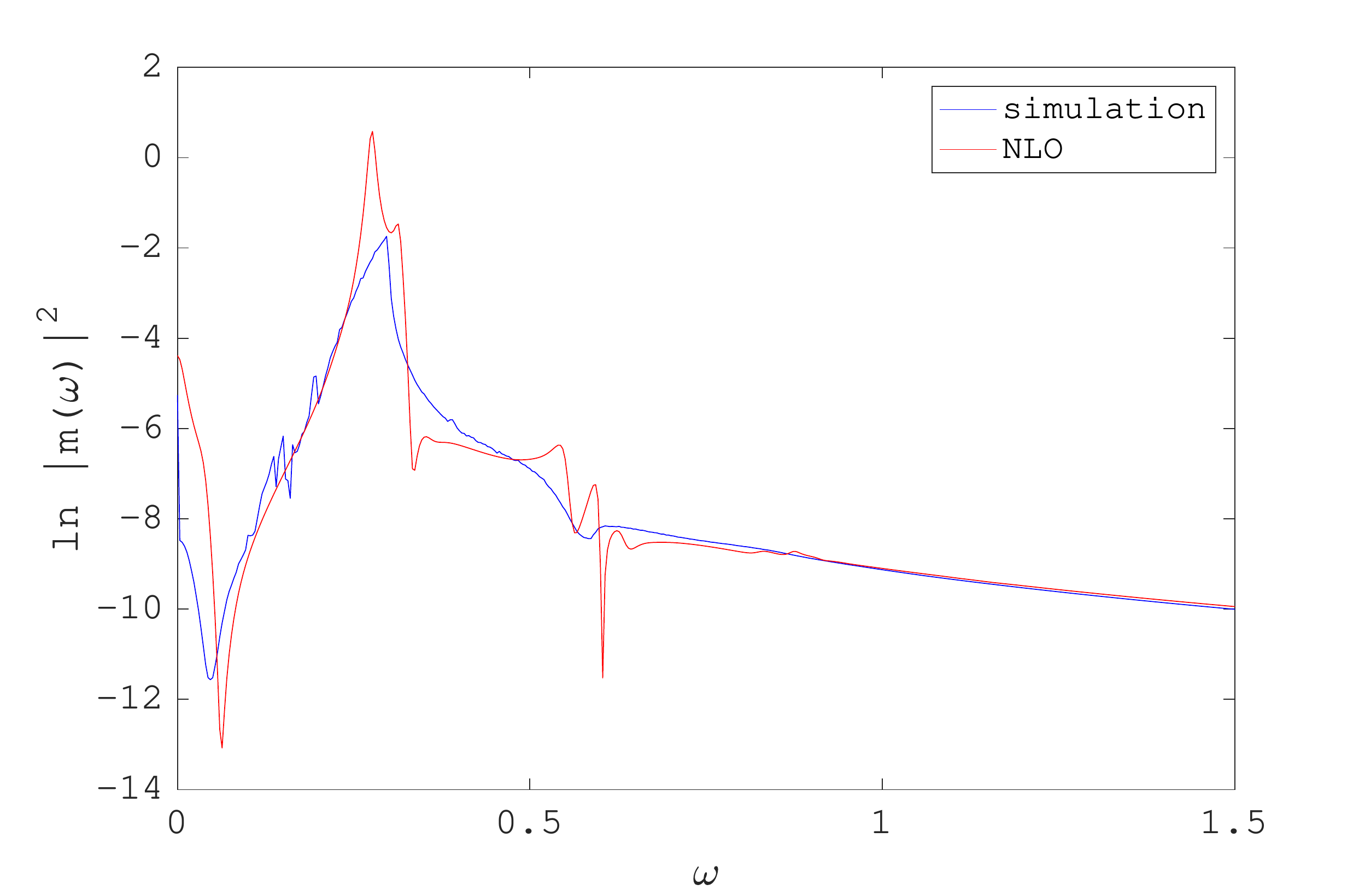}
\caption{Comparison between the Fourier power spectra of the solution of Eqs.\ \eqref{eq:Idot_NLO_diss}, with $\gamma_{(2)} = 0.1$ and $\gamma_{(4)} = 0.05$ (red line), and of the $m(t)$ obtained with a numerical simulation of an HMF model with $N = 2\times 10^6$ particles (blue line). Here $m_0 = 0.775$ and $b = 3.38$. The vertical scale is logarithmic.}
\label{fig:fourier_NLO}  
\end{figure}

Finally, we expect that going to the next-to-leading order should improve also the description of the actual time evolution $m(t)$, because finer details on the dynamics are now taken into account. Indeed, Fig.\ \ref{fig:fourier_NLO} shows that for the same value of $m_0$ considered in Fig.\ \ref{fig:fourier_LO} the Fourier power spectrum of $m(t)$ obtained by means of a numerical simulation of the HMF model is better reproduced by the solution of Eqs.\ \eqref{eq:Idot_NLO_diss} than by the leading-order solution. 

\section{Concluding remarks}
\label{sec_conclusions}

We have shown that an effective theory of collisionless (violent) relaxation after a cold collapse is possible, at least in the case of a very simple toy model with long-range interactions, the HMF model. Exploiting the fact that during collisionless relaxation the dynamics relocates of finer and finer scales, a description of the evolution of the collective properties of the system, blind to the fine structure, can be done by looking at the lowest-order moments of the distribution function. To the leading order we have shown that a description of the virial oscillations (without damping) naturally emerges as equivalent to the motion of a fictive particle in a one-dimensional potential. Then, inserting the contribution of the higher-order moments in an effective way, we have explicitly derived a dissipative equation describing virial oscillations and their damping, including a renormalization of the effective potential and yielding predictions for the value of the magnetization $m$ after the damping in very good agreement with numerical simulations. The derivation of the effective dissipation is inspired by Caldeira and Leggett's treatment of open quantum systems \cite{CaldeiraLeggett:prl1981,CaldeiraLeggett:physa1983}. As already mentioned in Sec.\ \ref{subsec_dissipation}, Hagstrom and Morrison \cite{HagstromMorrison:physicad2011} already discussed a relation between collisionless relaxation and the Caldeira-Leggett model: in the linear and homogeneous case they found an explicit mapping between the Vlasov-Poisson system and the Hamilton equations of the Caldeira-Leggett model. Although finding an analogous transformation valid in the fully nonlinear and non-homogeneous case is surely a very difficult task, it would put our  approach on a much more rigorous basis, so that it is a line that is worth pursuing. Work is in progress along this direction. 

In this paper we have considered only completely cold initial conditions, to which our theoretical approach seems best suited. However, with some small modifications this approach proves useful and effective also beyond the cold collapse: the extension to generic initial conditions for the HMF model will be presented in a forthcoming paper  \cite{violent_rel_II}. Work is also in progress on going beyond the HMF model, extending the approach to more complicated systems.

Apart from the possibility of making predictions on the quasi-stationary state reached after a cold collapse in the HMF model, the approach we have presented here can be considered a first, preliminary step towards a theoretical approach to collisionless relaxation that fully exploits the hierarchy of scales that is produced during the Vlasov evolution in order to derive an effective dynamical evolution, valid on sufficiently large scales, able to elucidate the relation between initial conditions and quasi-stationary states. We are admittedly still far from a comprehensive theory of violent relaxation predicting the shape of galaxies, yet the approach we have described here might give some hints towards this goal. Finally, it is worth recalling that the quasi-stationary state reached after the violent relaxation might not be the whole story, as it has been recently shown by means of numerical simulations of self-gravitating particles that structures like spiral arms and rings very similar to those observed in real galaxies may appear as long-living transients during the violent relaxation itself, when the initial conditions are not completely symmetric and the initial angular momentum does not vanish \cite{BenhaiemJoyceSylosLabini:apj2017}. This is a further indication that a deeper theoretical understanding of violent relaxation in long-range-interacting system is needed and could prove fruitful.

\bibliography{/Users/casetti/Work/Scripta/papers/bib/mybiblio,/Users/casetti/Work/Scripta/papers/bib/statmech,/Users/casetti/Work/Scripta/papers/bib/astro}

\begin{thebibliography}{38}%
\makeatletter
\providecommand \@ifxundefined [1]{%
 \@ifx{#1\undefined}
}%
\providecommand \@ifnum [1]{%
 \ifnum #1\expandafter \@firstoftwo
 \else \expandafter \@secondoftwo
 \fi
}%
\providecommand \@ifx [1]{%
 \ifx #1\expandafter \@firstoftwo
 \else \expandafter \@secondoftwo
 \fi
}%
\providecommand \natexlab [1]{#1}%
\providecommand \enquote  [1]{``#1''}%
\providecommand \bibnamefont  [1]{#1}%
\providecommand \bibfnamefont [1]{#1}%
\providecommand \citenamefont [1]{#1}%
\providecommand \href@noop [0]{\@secondoftwo}%
\providecommand \href [0]{\begingroup \@sanitize@url \@href}%
\providecommand \@href[1]{\@@startlink{#1}\@@href}%
\providecommand \@@href[1]{\endgroup#1\@@endlink}%
\providecommand \@sanitize@url [0]{\catcode `\\12\catcode `\$12\catcode
  `\&12\catcode `\#12\catcode `\^12\catcode `\_12\catcode `\%12\relax}%
\providecommand \@@startlink[1]{}%
\providecommand \@@endlink[0]{}%
\providecommand \url  [0]{\begingroup\@sanitize@url \@url }%
\providecommand \@url [1]{\endgroup\@href {#1}{\urlprefix }}%
\providecommand \urlprefix  [0]{URL }%
\providecommand \Eprint [0]{\href }%
\providecommand \doibase [0]{http://dx.doi.org/}%
\providecommand \selectlanguage [0]{\@gobble}%
\providecommand \bibinfo  [0]{\@secondoftwo}%
\providecommand \bibfield  [0]{\@secondoftwo}%
\providecommand \translation [1]{[#1]}%
\providecommand \BibitemOpen [0]{}%
\providecommand \bibitemStop [0]{}%
\providecommand \bibitemNoStop [0]{.\EOS\space}%
\providecommand \EOS [0]{\spacefactor3000\relax}%
\providecommand \BibitemShut  [1]{\csname bibitem#1\endcsname}%
\let\auto@bib@innerbib\@empty
\bibitem [{\citenamefont {Campa}\ \emph {et~al.}(2014)\citenamefont {Campa},
  \citenamefont {Dauxois}, \citenamefont {Fanelli},\ and\ \citenamefont
  {Ruffo}}]{CampaEtAl:book}%
  \BibitemOpen
  \bibfield  {author} {\bibinfo {author} {\bibfnamefont {A.}~\bibnamefont
  {Campa}}, \bibinfo {author} {\bibfnamefont {T.}~\bibnamefont {Dauxois}},
  \bibinfo {author} {\bibfnamefont {D.}~\bibnamefont {Fanelli}}, \ and\
  \bibinfo {author} {\bibfnamefont {S.}~\bibnamefont {Ruffo}},\ }\href@noop {}
  {\emph {\bibinfo {title} {Physics of Long-Range Interacting Systems}}}\
  (\bibinfo  {publisher} {Oxford University Press},\ \bibinfo {address}
  {Oxford},\ \bibinfo {year} {2014})\BibitemShut {NoStop}%
\bibitem [{\citenamefont {Campa}\ \emph {et~al.}(2009)\citenamefont {Campa},
  \citenamefont {Dauxois},\ and\ \citenamefont {Ruffo}}]{CampaEtAl:physrep}%
  \BibitemOpen
  \bibfield  {author} {\bibinfo {author} {\bibfnamefont {A.}~\bibnamefont
  {Campa}}, \bibinfo {author} {\bibfnamefont {T.}~\bibnamefont {Dauxois}}, \
  and\ \bibinfo {author} {\bibfnamefont {S.}~\bibnamefont {Ruffo}},\ }\href
  {\doibase 10.1016/j.physrep.2009.07.001} {\bibfield  {journal} {\bibinfo
  {journal} {Physics Reports}\ }\textbf {\bibinfo {volume} {480}},\ \bibinfo
  {pages} {57} (\bibinfo {year} {2009})}\BibitemShut {NoStop}%
\bibitem [{\citenamefont {Binney}\ and\ \citenamefont
  {Tremaine}(2008)}]{BinneyTremaine:book}%
  \BibitemOpen
  \bibfield  {author} {\bibinfo {author} {\bibfnamefont {J.}~\bibnamefont
  {Binney}}\ and\ \bibinfo {author} {\bibfnamefont {S.}~\bibnamefont
  {Tremaine}},\ }\href@noop {} {\emph {\bibinfo {title} {Galactic Dynamics}}},\
  \bibinfo {edition} {2nd}\ ed.\ (\bibinfo  {publisher} {Princeton University
  Press},\ \bibinfo {address} {Princeton},\ \bibinfo {year} {2008})\BibitemShut
  {NoStop}%
\bibitem [{\citenamefont {Nicholson}(1983)}]{Nicholson:book}%
  \BibitemOpen
  \bibfield  {author} {\bibinfo {author} {\bibfnamefont {D.~R.}\ \bibnamefont
  {Nicholson}},\ }\href@noop {} {\emph {\bibinfo {title} {Introduction to
  Plasma Theory}}}\ (\bibinfo  {publisher} {Wiley},\ \bibinfo {address} {New
  York},\ \bibinfo {year} {1983})\BibitemShut {NoStop}%
\bibitem [{\citenamefont {Sch\"utz}\ and\ \citenamefont
  {Morigi}(2014)}]{SchutzMorigi:prl2014}%
  \BibitemOpen
  \bibfield  {author} {\bibinfo {author} {\bibfnamefont {S.}~\bibnamefont
  {Sch\"utz}}\ and\ \bibinfo {author} {\bibfnamefont {G.}~\bibnamefont
  {Morigi}},\ }\href {\doibase 10.1103/PhysRevLett.113.203002} {\bibfield
  {journal} {\bibinfo  {journal} {Phys. Rev. Lett.}\ }\textbf {\bibinfo
  {volume} {113}},\ \bibinfo {pages} {203002} (\bibinfo {year}
  {2014})}\BibitemShut {NoStop}%
\bibitem [{\citenamefont {{Gupta}}\ and\ \citenamefont
  {{Casetti}}(2016)}]{njp2016}%
  \BibitemOpen
  \bibfield  {author} {\bibinfo {author} {\bibfnamefont {S.}~\bibnamefont
  {{Gupta}}}\ and\ \bibinfo {author} {\bibfnamefont {L.}~\bibnamefont
  {{Casetti}}},\ }\href {\doibase 10.1088/1367-2630/18/10/103051} {\bibfield
  {journal} {\bibinfo  {journal} {New Journal of Physics}\ }\textbf {\bibinfo
  {volume} {18}},\ \bibinfo {eid} {103051} (\bibinfo {year}
  {2016})}\BibitemShut {NoStop}%
\bibitem [{\citenamefont {Lynden-Bell}(1967)}]{Lynden-Bell:mnras1967}%
  \BibitemOpen
  \bibfield  {author} {\bibinfo {author} {\bibfnamefont {D.}~\bibnamefont
  {Lynden-Bell}},\ }\href {\doibase 10.1093/mnras/136.1.101} {\bibfield
  {journal} {\bibinfo  {journal} {Mon. Not. Royal Astron. Soc.}\ }\textbf
  {\bibinfo {volume} {136}},\ \bibinfo {pages} {101} (\bibinfo {year}
  {1967})}\BibitemShut {NoStop}%
\bibitem [{\citenamefont {{H{\'e}non}}(1964)}]{Henon:1964AnnAp1964}%
  \BibitemOpen
  \bibfield  {author} {\bibinfo {author} {\bibfnamefont {M.}~\bibnamefont
  {{H{\'e}non}}},\ }\href@noop {} {\bibfield  {journal} {\bibinfo  {journal}
  {Annales d'Astrophysique}\ }\textbf {\bibinfo {volume} {27}},\ \bibinfo
  {pages} {83} (\bibinfo {year} {1964})}\BibitemShut {NoStop}%
\bibitem [{\citenamefont {{van Albada}}(1982)}]{vanAlbada:mnras1982}%
  \BibitemOpen
  \bibfield  {author} {\bibinfo {author} {\bibfnamefont {T.~S.}\ \bibnamefont
  {{van Albada}}},\ }\href {\doibase 10.1093/mnras/201.4.939} {\bibfield
  {journal} {\bibinfo  {journal} {Mon. Not. Royal Astron. Soc.}\ }\textbf
  {\bibinfo {volume} {201}},\ \bibinfo {pages} {939} (\bibinfo {year}
  {1982})}\BibitemShut {NoStop}%
\bibitem [{\citenamefont {Sylos~Labini}(2012)}]{Sylos:mnras2012}%
  \BibitemOpen
  \bibfield  {author} {\bibinfo {author} {\bibfnamefont {F.}~\bibnamefont
  {Sylos~Labini}},\ }\href {\doibase 10.1111/j.1365-2966.2012.21019.x}
  {\bibfield  {journal} {\bibinfo  {journal} {Mon. Not. Royal Astron. Soc.}\
  }\textbf {\bibinfo {volume} {423}},\ \bibinfo {pages} {1610} (\bibinfo {year}
  {2012})}\BibitemShut {NoStop}%
\bibitem [{\citenamefont {{Di Cintio}}\ \emph {et~al.}(2018)\citenamefont {{Di
  Cintio}}, \citenamefont {{Gupta}},\ and\ \citenamefont
  {{Casetti}}}]{mnras2018}%
  \BibitemOpen
  \bibfield  {author} {\bibinfo {author} {\bibfnamefont {P.}~\bibnamefont {{Di
  Cintio}}}, \bibinfo {author} {\bibfnamefont {S.}~\bibnamefont {{Gupta}}}, \
  and\ \bibinfo {author} {\bibfnamefont {L.}~\bibnamefont {{Casetti}}},\ }\href
  {\doibase 10.1093/mnras/stx3244} {\bibfield  {journal} {\bibinfo  {journal}
  {Mon. Not. Royal Astron. Soc.}\ }\textbf {\bibinfo {volume} {475}},\ \bibinfo
  {pages} {1137} (\bibinfo {year} {2018})}\BibitemShut {NoStop}%
\bibitem [{\citenamefont {Teles}\ \emph {et~al.}(2015)\citenamefont {Teles},
  \citenamefont {Gupta}, \citenamefont {Di~Cintio},\ and\ \citenamefont
  {Casetti}}]{prerap2015}%
  \BibitemOpen
  \bibfield  {author} {\bibinfo {author} {\bibfnamefont {T.~N.}\ \bibnamefont
  {Teles}}, \bibinfo {author} {\bibfnamefont {S.}~\bibnamefont {Gupta}},
  \bibinfo {author} {\bibfnamefont {P.}~\bibnamefont {Di~Cintio}}, \ and\
  \bibinfo {author} {\bibfnamefont {L.}~\bibnamefont {Casetti}},\ }\href
  {\doibase 10.1103/PhysRevE.92.020101} {\bibfield  {journal} {\bibinfo
  {journal} {Phys. Rev. E}\ }\textbf {\bibinfo {volume} {92}},\ \bibinfo
  {pages} {020101} (\bibinfo {year} {2015})}\BibitemShut {NoStop}%
\bibitem [{\citenamefont {Bertin}(2000)}]{Bertin:book}%
  \BibitemOpen
  \bibfield  {author} {\bibinfo {author} {\bibfnamefont {G.}~\bibnamefont
  {Bertin}},\ }\href@noop {} {\emph {\bibinfo {title} {Dynamics of Galaxies}}}\
  (\bibinfo  {publisher} {Cambridge University Press},\ \bibinfo {address}
  {Cambridge},\ \bibinfo {year} {2000})\BibitemShut {NoStop}%
\bibitem [{\citenamefont {Levin}\ \emph {et~al.}(2014)\citenamefont {Levin},
  \citenamefont {Pakter}, \citenamefont {Rizzato}, \citenamefont {Teles},\ and\
  \citenamefont {Benetti}}]{LevinEtAlphysrep:2014}%
  \BibitemOpen
  \bibfield  {author} {\bibinfo {author} {\bibfnamefont {Y.}~\bibnamefont
  {Levin}}, \bibinfo {author} {\bibfnamefont {R.}~\bibnamefont {Pakter}},
  \bibinfo {author} {\bibfnamefont {F.~B.}\ \bibnamefont {Rizzato}}, \bibinfo
  {author} {\bibfnamefont {T.~N.}\ \bibnamefont {Teles}}, \ and\ \bibinfo
  {author} {\bibfnamefont {F.~P.~C.}\ \bibnamefont {Benetti}},\ }\href
  {\doibase http://dx.doi.org/10.1016/j.physrep.2013.10.001} {\bibfield
  {journal} {\bibinfo  {journal} {Physics Reports}\ }\textbf {\bibinfo {volume}
  {535}},\ \bibinfo {pages} {1 } (\bibinfo {year} {2014})}\BibitemShut
  {NoStop}%
\bibitem [{\citenamefont {{Leoncini}}\ \emph {et~al.}(2009)\citenamefont
  {{Leoncini}}, \citenamefont {{Van Den Berg}},\ and\ \citenamefont
  {{Fanelli}}}]{LeonciniVanDenBergFanelli:epl2009}%
  \BibitemOpen
  \bibfield  {author} {\bibinfo {author} {\bibfnamefont {X.}~\bibnamefont
  {{Leoncini}}}, \bibinfo {author} {\bibfnamefont {T.~L.}\ \bibnamefont {{Van
  Den Berg}}}, \ and\ \bibinfo {author} {\bibfnamefont {D.}~\bibnamefont
  {{Fanelli}}},\ }\href {\doibase 10.1209/0295-5075/86/20002} {\bibfield
  {journal} {\bibinfo  {journal} {EPL (Europhysics Letters)}\ }\textbf
  {\bibinfo {volume} {86}},\ \bibinfo {pages} {20002} (\bibinfo {year}
  {2009})}\BibitemShut {NoStop}%
\bibitem [{\citenamefont {Campa}\ and\ \citenamefont
  {Chavanis}(2010)}]{CampaChavanis:jstat2010}%
  \BibitemOpen
  \bibfield  {author} {\bibinfo {author} {\bibfnamefont {A.}~\bibnamefont
  {Campa}}\ and\ \bibinfo {author} {\bibfnamefont {P.-H.}\ \bibnamefont
  {Chavanis}},\ }\href {\doibase 10.1088/1742-5468/2010/06/P06001} {\bibfield
  {journal} {\bibinfo  {journal} {Journal of Statistical Mechanics: Theory and
  Experiment}\ }\textbf {\bibinfo {volume} {2010}},\ \bibinfo {pages} {P06001}
  (\bibinfo {year} {2010})}\BibitemShut {NoStop}%
\bibitem [{\citenamefont {Chavanis}\ and\ \citenamefont
  {Campa}(2010)}]{ChavanisCampa:epjb2010}%
  \BibitemOpen
  \bibfield  {author} {\bibinfo {author} {\bibfnamefont {P.-H.}\ \bibnamefont
  {Chavanis}}\ and\ \bibinfo {author} {\bibfnamefont {A.}~\bibnamefont
  {Campa}},\ }\href {\doibase 10.1140/epjb/e2010-00243-x} {\bibfield  {journal}
  {\bibinfo  {journal} {Eur. Phys. J. B}\ }\textbf {\bibinfo {volume} {76}},\
  \bibinfo {pages} {581} (\bibinfo {year} {2010})}\BibitemShut {NoStop}%
\bibitem [{\citenamefont {{Assllani}}\ \emph {et~al.}(2012)\citenamefont
  {{Assllani}}, \citenamefont {{Fanelli}}, \citenamefont {{Turchi}},
  \citenamefont {{Carletti}},\ and\ \citenamefont
  {{Leoncini}}}]{AssllaniEtAl:pre2012}%
  \BibitemOpen
  \bibfield  {author} {\bibinfo {author} {\bibfnamefont {M.}~\bibnamefont
  {{Assllani}}}, \bibinfo {author} {\bibfnamefont {D.}~\bibnamefont
  {{Fanelli}}}, \bibinfo {author} {\bibfnamefont {A.}~\bibnamefont {{Turchi}}},
  \bibinfo {author} {\bibfnamefont {T.}~\bibnamefont {{Carletti}}}, \ and\
  \bibinfo {author} {\bibfnamefont {X.}~\bibnamefont {{Leoncini}}},\ }\href
  {\doibase 10.1103/PhysRevE.85.021148} {\bibfield  {journal} {\bibinfo
  {journal} {\pre}\ }\textbf {\bibinfo {volume} {85}},\ \bibinfo {eid} {021148}
  (\bibinfo {year} {2012})}\BibitemShut {NoStop}%
\bibitem [{\citenamefont {Benetti}\ \emph {et~al.}(2014)\citenamefont
  {Benetti}, \citenamefont {Ribeiro-Teixeira}, \citenamefont {Pakter},\ and\
  \citenamefont {Levin}}]{BenettiEtAl:prl2014}%
  \BibitemOpen
  \bibfield  {author} {\bibinfo {author} {\bibfnamefont {F.~P.~C.}\
  \bibnamefont {Benetti}}, \bibinfo {author} {\bibfnamefont {A.~C.}\
  \bibnamefont {Ribeiro-Teixeira}}, \bibinfo {author} {\bibfnamefont
  {R.}~\bibnamefont {Pakter}}, \ and\ \bibinfo {author} {\bibfnamefont
  {Y.}~\bibnamefont {Levin}},\ }\href {\doibase 10.1103/PhysRevLett.113.100602}
  {\bibfield  {journal} {\bibinfo  {journal} {Phys. Rev. Lett.}\ }\textbf
  {\bibinfo {volume} {113}},\ \bibinfo {pages} {100602} (\bibinfo {year}
  {2014})}\BibitemShut {NoStop}%
\bibitem [{\citenamefont {{Kandrup}}(1998)}]{Kandrup:apj1998}%
  \BibitemOpen
  \bibfield  {author} {\bibinfo {author} {\bibfnamefont {H.~E.}\ \bibnamefont
  {{Kandrup}}},\ }\href {\doibase 10.1086/305721} {\bibfield  {journal}
  {\bibinfo  {journal} {\apj}\ }\textbf {\bibinfo {volume} {500}},\ \bibinfo
  {pages} {120} (\bibinfo {year} {1998})}\BibitemShut {NoStop}%
\bibitem [{\citenamefont {Barr\'{e}}\ \emph {et~al.}(2010)\citenamefont
  {Barr\'{e}}, \citenamefont {Olivetti},\ and\ \citenamefont
  {Yamaguchi}}]{BarreOlivettiYamaguchi:jstat2010}%
  \BibitemOpen
  \bibfield  {author} {\bibinfo {author} {\bibfnamefont {J.}~\bibnamefont
  {Barr\'{e}}}, \bibinfo {author} {\bibfnamefont {A.}~\bibnamefont {Olivetti}},
  \ and\ \bibinfo {author} {\bibfnamefont {Y.~Y.}\ \bibnamefont {Yamaguchi}},\
  }\href {\doibase 10.1088/1751-8113/44/40/405502} {\bibfield  {journal}
  {\bibinfo  {journal} {Journal of Statistical Mechanics: Theory and
  Experiment}\ }\textbf {\bibinfo {volume} {2010}},\ \bibinfo {pages} {P08002}
  (\bibinfo {year} {2010})}\BibitemShut {NoStop}%
\bibitem [{\citenamefont {Barr\'{e}}\ \emph {et~al.}(2011)\citenamefont
  {Barr\'{e}}, \citenamefont {Olivetti},\ and\ \citenamefont
  {Yamaguchi}}]{BarreOlivettiYamaguchi:jphysa2011}%
  \BibitemOpen
  \bibfield  {author} {\bibinfo {author} {\bibfnamefont {J.}~\bibnamefont
  {Barr\'{e}}}, \bibinfo {author} {\bibfnamefont {A.}~\bibnamefont {Olivetti}},
  \ and\ \bibinfo {author} {\bibfnamefont {Y.~Y.}\ \bibnamefont {Yamaguchi}},\
  }\href {\doibase 10.1088/1742-5468/2010/08/P08002} {\bibfield  {journal}
  {\bibinfo  {journal} {Journal of Physics A: Mathematical and Theoretical}\
  }\textbf {\bibinfo {volume} {44}},\ \bibinfo {pages} {405502} (\bibinfo
  {year} {2011})}\BibitemShut {NoStop}%
\bibitem [{\citenamefont {{Tremaine}}\ \emph {et~al.}(1986)\citenamefont
  {{Tremaine}}, \citenamefont {{H\'enon}},\ and\ \citenamefont
  {{Lynden-Bell}}}]{TremaineHenonLynden-Bell:mnras1986}%
  \BibitemOpen
  \bibfield  {author} {\bibinfo {author} {\bibfnamefont {S.}~\bibnamefont
  {{Tremaine}}}, \bibinfo {author} {\bibfnamefont {M.}~\bibnamefont
  {{H\'enon}}}, \ and\ \bibinfo {author} {\bibfnamefont {D.}~\bibnamefont
  {{Lynden-Bell}}},\ }\href {\doibase 10.1093/mnras/219.2.285} {\bibfield
  {journal} {\bibinfo  {journal} {Mon. Not. Royal Astron. Soc.}\ }\textbf
  {\bibinfo {volume} {219}},\ \bibinfo {pages} {285} (\bibinfo {year}
  {1986})}\BibitemShut {NoStop}%
\bibitem [{\citenamefont {Mohout}\ and\ \citenamefont
  {Villani}(2011)}]{MohoutVillani:actamath2011}%
  \BibitemOpen
  \bibfield  {author} {\bibinfo {author} {\bibfnamefont {C.}~\bibnamefont
  {Mohout}}\ and\ \bibinfo {author} {\bibfnamefont {C.}~\bibnamefont
  {Villani}},\ }\href {\doibase 10.1007/s11511-011-0068-9} {\bibfield
  {journal} {\bibinfo  {journal} {Acta Mathematica}\ }\textbf {\bibinfo
  {volume} {207}},\ \bibinfo {pages} {29} (\bibinfo {year} {2011})}\BibitemShut
  {NoStop}%
\bibitem [{\citenamefont {Caldeira}\ and\ \citenamefont
  {Leggett}(1981)}]{CaldeiraLeggett:prl1981}%
  \BibitemOpen
  \bibfield  {author} {\bibinfo {author} {\bibfnamefont {A.~O.}\ \bibnamefont
  {Caldeira}}\ and\ \bibinfo {author} {\bibfnamefont {A.~J.}\ \bibnamefont
  {Leggett}},\ }\href {\doibase 10.1103/PhysRevLett.46.211} {\bibfield
  {journal} {\bibinfo  {journal} {Phys. Rev. Lett.}\ }\textbf {\bibinfo
  {volume} {46}},\ \bibinfo {pages} {211} (\bibinfo {year} {1981})}\BibitemShut
  {NoStop}%
\bibitem [{\citenamefont {Caldeira}\ and\ \citenamefont
  {Leggett}(1983)}]{CaldeiraLeggett:physa1983}%
  \BibitemOpen
  \bibfield  {author} {\bibinfo {author} {\bibfnamefont {A.~O.}\ \bibnamefont
  {Caldeira}}\ and\ \bibinfo {author} {\bibfnamefont {A.~J.}\ \bibnamefont
  {Leggett}},\ }\href {\doibase 10.1016/0378-4371(83)90013-4} {\bibfield
  {journal} {\bibinfo  {journal} {Physica A}\ }\textbf {\bibinfo {volume}
  {121}},\ \bibinfo {pages} {587} (\bibinfo {year} {1983})}\BibitemShut
  {NoStop}%
\bibitem [{\citenamefont {Giachetti}\ \emph {et~al.}(2019)\citenamefont
  {Giachetti}, \citenamefont {Santini},\ and\ \citenamefont
  {Casetti}}]{violent_rel_II}%
  \BibitemOpen
  \bibfield  {author} {\bibinfo {author} {\bibfnamefont {G.}~\bibnamefont
  {Giachetti}}, \bibinfo {author} {\bibfnamefont {A.}~\bibnamefont {Santini}},
  \ and\ \bibinfo {author} {\bibfnamefont {L.}~\bibnamefont {Casetti}},\
  }\href@noop {} {\  (\bibinfo {year} {2019})},\ \bibinfo {note} {in
  preparation}\BibitemShut {NoStop}%
\bibitem [{\citenamefont {Messer}\ and\ \citenamefont
  {Spohn}(1982)}]{MesserSpohn:jsp1982}%
  \BibitemOpen
  \bibfield  {author} {\bibinfo {author} {\bibfnamefont {J.}~\bibnamefont
  {Messer}}\ and\ \bibinfo {author} {\bibfnamefont {H.}~\bibnamefont {Spohn}},\
  }\href@noop {} {\bibfield  {journal} {\bibinfo  {journal} {J. Stat. Phys.}\
  }\textbf {\bibinfo {volume} {29}},\ \bibinfo {pages} {561} (\bibinfo {year}
  {1982})}\BibitemShut {NoStop}%
\bibitem [{\citenamefont {Battle}(1977)}]{Battle:cmp1977}%
  \BibitemOpen
  \bibfield  {author} {\bibinfo {author} {\bibfnamefont {G.~A.}\ \bibnamefont
  {Battle}},\ }\href@noop {} {\bibfield  {journal} {\bibinfo  {journal} {Comm.
  Math. Phys.}\ }\textbf {\bibinfo {volume} {55}},\ \bibinfo {pages} {229}
  (\bibinfo {year} {1977})}\BibitemShut {NoStop}%
\bibitem [{\citenamefont {Antoni}\ and\ \citenamefont
  {Ruffo}(1995)}]{AntoniRuffo:pre1995}%
  \BibitemOpen
  \bibfield  {author} {\bibinfo {author} {\bibfnamefont {M.}~\bibnamefont
  {Antoni}}\ and\ \bibinfo {author} {\bibfnamefont {S.}~\bibnamefont {Ruffo}},\
  }\href {\doibase 10.1103/PhysRevE.52.2361} {\bibfield  {journal} {\bibinfo
  {journal} {Phys. Rev. E}\ }\textbf {\bibinfo {volume} {52}},\ \bibinfo
  {pages} {2361} (\bibinfo {year} {1995})}\BibitemShut {NoStop}%
\bibitem [{\citenamefont {Tatekawa}\ \emph {et~al.}(2005)\citenamefont
  {Tatekawa}, \citenamefont {Bouchet}, \citenamefont {Dauxois},\ and\
  \citenamefont {Ruffo}}]{TatekawaEtAl:pre2005}%
  \BibitemOpen
  \bibfield  {author} {\bibinfo {author} {\bibfnamefont {T.}~\bibnamefont
  {Tatekawa}}, \bibinfo {author} {\bibfnamefont {F.}~\bibnamefont {Bouchet}},
  \bibinfo {author} {\bibfnamefont {T.}~\bibnamefont {Dauxois}}, \ and\
  \bibinfo {author} {\bibfnamefont {S.}~\bibnamefont {Ruffo}},\ }\href
  {\doibase 10.1103/PhysRevE.71.056111} {\bibfield  {journal} {\bibinfo
  {journal} {Phys. Rev. E}\ }\textbf {\bibinfo {volume} {71}},\ \bibinfo
  {pages} {056111} (\bibinfo {year} {2005})}\BibitemShut {NoStop}%
\bibitem [{\citenamefont {Pichon}(1994)}]{Pichon:thesis}%
  \BibitemOpen
  \bibfield  {author} {\bibinfo {author} {\bibfnamefont {C.}~\bibnamefont
  {Pichon}},\ }\href@noop {} {Ph.D. thesis},\ \bibinfo  {school} {Cambridge
  University} (\bibinfo {year} {1994})\BibitemShut {NoStop}%
\bibitem [{\citenamefont {Fouvry}\ and\ \citenamefont
  {Bar-Or}(2018)}]{FouvryBarOr:mnras2018}%
  \BibitemOpen
  \bibfield  {author} {\bibinfo {author} {\bibfnamefont {J.-B.}\ \bibnamefont
  {Fouvry}}\ and\ \bibinfo {author} {\bibfnamefont {B.}~\bibnamefont
  {Bar-Or}},\ }\href {\doibase 10.1093/mnras/sty2571} {\bibfield  {journal}
  {\bibinfo  {journal} {Mon. Not. Royal Astron. Soc.}\ }\textbf {\bibinfo
  {volume} {481}},\ \bibinfo {pages} {4566} (\bibinfo {year}
  {2018})}\BibitemShut {NoStop}%
\bibitem [{\citenamefont {Casetti}(1995)}]{physscr1995}%
  \BibitemOpen
  \bibfield  {author} {\bibinfo {author} {\bibfnamefont {L.}~\bibnamefont
  {Casetti}},\ }\href {\doibase 10.1088/0031-8949/51/1/005} {\bibfield
  {journal} {\bibinfo  {journal} {Physica Scripta}\ }\textbf {\bibinfo {volume}
  {51}},\ \bibinfo {pages} {29} (\bibinfo {year} {1995})}\BibitemShut {NoStop}%
\bibitem [{\citenamefont {Villain}(1975)}]{Villain:jphys1975}%
  \BibitemOpen
  \bibfield  {author} {\bibinfo {author} {\bibfnamefont {J.}~\bibnamefont
  {Villain}},\ }\href {\doibase 10.1051/jphys:01975003606058100} {\bibfield
  {journal} {\bibinfo  {journal} {Journal de Physique}\ }\textbf {\bibinfo
  {volume} {36}},\ \bibinfo {pages} {581} (\bibinfo {year} {1975})}\BibitemShut
  {NoStop}%
\bibitem [{\citenamefont {Zwanzig}(2001)}]{Zwanzig:book}%
  \BibitemOpen
  \bibfield  {author} {\bibinfo {author} {\bibfnamefont {R.}~\bibnamefont
  {Zwanzig}},\ }\href@noop {} {\emph {\bibinfo {title} {Nonequilibrium
  statistical mechanics}}}\ (\bibinfo  {publisher} {Oxford},\ \bibinfo
  {address} {New York},\ \bibinfo {year} {2001})\BibitemShut {NoStop}%
\bibitem [{\citenamefont {Hagstrom}\ and\ \citenamefont
  {Morrison}(2011)}]{HagstromMorrison:physicad2011}%
  \BibitemOpen
  \bibfield  {author} {\bibinfo {author} {\bibfnamefont {G.~I.}\ \bibnamefont
  {Hagstrom}}\ and\ \bibinfo {author} {\bibfnamefont {P.~J.}\ \bibnamefont
  {Morrison}},\ }\href {\doibase 10.1016/j.physd.2011.02.007} {\bibfield
  {journal} {\bibinfo  {journal} {Physica D}\ }\textbf {\bibinfo {volume}
  {240}},\ \bibinfo {pages} {1652} (\bibinfo {year} {2011})}\BibitemShut
  {NoStop}%
\bibitem [{\citenamefont {Benhaiem}\ \emph {et~al.}(2017)\citenamefont
  {Benhaiem}, \citenamefont {Joyce},\ and\ \citenamefont {{Sylos
  Labini}}}]{BenhaiemJoyceSylosLabini:apj2017}%
  \BibitemOpen
  \bibfield  {author} {\bibinfo {author} {\bibfnamefont {D.}~\bibnamefont
  {Benhaiem}}, \bibinfo {author} {\bibfnamefont {M.}~\bibnamefont {Joyce}}, \
  and\ \bibinfo {author} {\bibfnamefont {F.}~\bibnamefont {{Sylos Labini}}},\
  }\href {\doibase 10.3847/1538-4357/aa96a7} {\bibfield  {journal} {\bibinfo
  {journal} {\apj}\ }\textbf {\bibinfo {volume} {851}},\ \bibinfo {pages} {19}
  (\bibinfo {year} {2017})}\BibitemShut {NoStop}%
\end{thebibliography}%

\appendix

\section{Virial theorem, thermal equilibrium and the leading-order prediction of the magnetization}
\label{app:virial}

As mentioned in Sec.\ \ref{subsec_virial}, the bare leading-order prediction $\overline{m}$ of the magnetization in the quasi-stationary state after a cold collapse given by Eq.\ \eqref{eq:mbar_LO} could have been found using the virial theorem. This follows from the fact that at the leading order the mean-field potential $U(\vartheta)$ is harmonic,
\beq
U(\vartheta) = -m\left(1 - \frac{\vartheta^2}{2} \right)~;
\eeq
applying the virial theorem to the quasi-stationary state we thus have
\beq
\langle p^2 \rangle =\left \langle \vartheta \frac{\partial U}{\partial \vartheta} \right\rangle = \overline{m} \left \langle \vartheta^2\right\rangle ~,
\eeq 
and using $\overline{m} = 1- \langle\vartheta^2\rangle/2$ and the conservation of energy $\varepsilon = \langle p^2 \rangle/2 - \overline{m}^2/2$ we can write 
\beq
2\varepsilon + \overline{m}^2 = 2\overline{m}(1- \overline{m})~.
\label{eq:eps_mbar}
\eeq
In a cold collapse $\varepsilon = - m_0^2/2$, and inserting the latter into Eq.\ \eqref{eq:eps_mbar} we have
\beq
3\overline{m}^2 - 2 \overline{m} - m_0 =0\,,
\eeq
whose only positive solution is
\beq
\overline{m} = \frac{1+\sqrt{1 + 3m_0^2}}{3}~,
\eeq
that is, Eq.\ \eqref{eq:mbar_LO}.

\begin{figure}
\includegraphics[width = 0.7\textwidth]{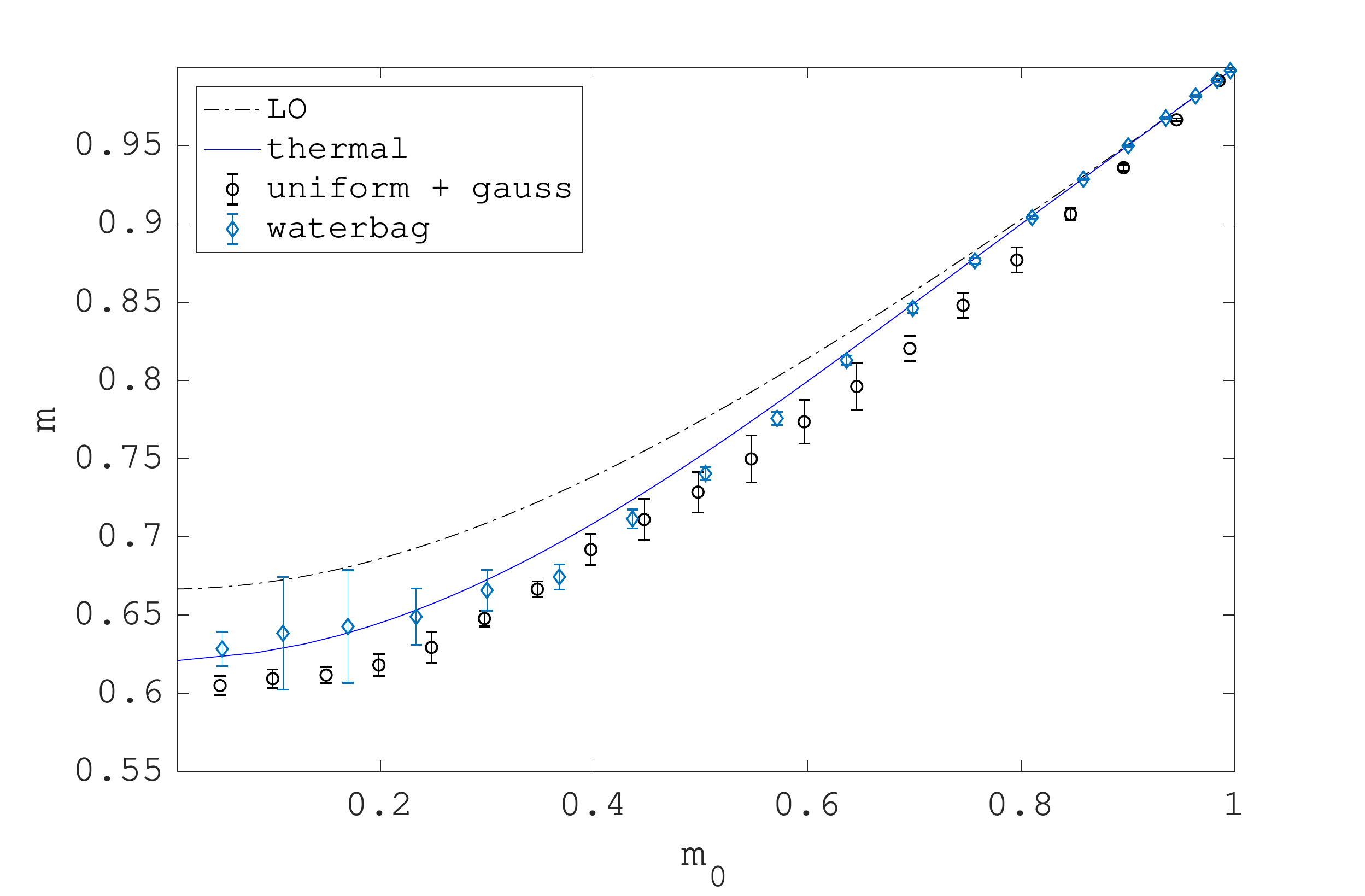}
\caption{Comparison between the bare leading-order theoretical predictions for the asymptotic magnetization after a cold collapse $\overline{m}$ given by Eq.\ \eqref{eq:mbar_LO} (black dash-dotted line), the magnetization in thermal equilibrium $m_{\text{eq}}$ given by Eqs.\ \eqref{eq:selfcons} and \eqref{eq:beta} (blue solid curve), and the results of numerical simulations already shown in Fig.\ \ref{fig:m}.}
\label{fig:thermal}  
\end{figure}

It is interesting to compare the latter result and the actual values of $\overline{m}$ found in numerical simulations with the magnetization in thermal equilibrium. As shown in \cite{AntoniRuffo:pre1995,CampaEtAl:book} the magnetization $m_{\text{eq}}$ in thermal equilibrium is implicitly given as a function of the total energy $\varepsilon$ by the solution of the equation
\beq
m_{\text{eq}} = \frac{I_1\left(\beta m_{\text{eq}}\right)}{I_0\left(\beta m_{\text{eq}}\right)}\,,
\label{eq:selfcons}
\eeq
where $I_k(x)$ is the modified Bessel function of order $k$ and $\beta$ is defined by
\beq
\varepsilon = \frac{1}{2\beta} - \frac{m^2_{\text{eq}}}{2}~.  
\label{eq:beta}
\eeq
The magnetization in thermal equilibrium after a cold collapse as a function of the initial magnetization, $m_{\text{eq}}(m_0)$, is thus obtained imposing $\varepsilon = -m_0^2/2$ in Eq.\ \eqref{eq:beta}. The thermal equilibrium prediction is plotted in Fig.\ \ref{fig:thermal} together with the leading-order prediction $\overline{m}$ given by Eq.\ \eqref{eq:mbar_LO} and compared to the simulation data. It is apparent that the numerical results obtained starting from initial conditions where a Gaussian overdensity is superimposed on a uniform background yield magnetizations in the quasi-stationary state after the cold collapse that are definitely different from the thermal predictions, although the difference is not very large, while cold waterbag initial conditions lead to quasi-stationary states where the magnetization might be consistent with the thermal equilibrium value, at least for sufficiently high or sufficiently low initial magnetizations. This does not mean, however, that such quasi-stationary states are close to thermal: it only means that the differences between the quasi-stationary state and the thermal state show up when we look at finer scales of the distribution function and are not apparent at the larger scale captured by $m$.

\section{Solution of the equations of motion of the auxiliary oscillators}
\label{app:calculations}

Here we describe how to solve the equations of motion of the auxiliary oscillators introduced in Sec.\ \ref{subsec_dissipation} to model the contribution of the higher-order moments to the dynamics of the low-order ones, i.e., Eqs.\ \eqref{eq:ddotq}, that we rewrite here for convenience,
\beq
\ddot q_k = - \omega_k^2 q_k + c_k g(x)\,,
\label{eq:ddotq_app}
\eeq
where $k = 1,\ldots,M$. 

The solution is a direct adaptation of that used to solve the analogous equations for the classical Caldeira-Leggett model, where $g(x) = x$, to the case of a generic function $g(x)$. We define the Laplace transform $\tilde f (s)$ of a generic function of time $f(t)$ as follows
\beq
\mathcal{L}(f) = \tilde f (s) = \int_0^\infty f(t)\, e^{-st}\, dt\,,
\eeq
so that
\beq
\mathcal{L}^{-1}(\tilde f) = f(t) = \frac{1}{2\pi i}\int_{\sigma - i\infty}^{\sigma + i\infty} \tilde f (s) \, 
e^{st}\, ds\,,
\label{eq:antilaplace}
\eeq
where $s\in \mathbb{R}$ is larger than the real part of any pole of $\tilde f(s)$. If we Laplace-transform Eqs.\ \eqref{eq:ddotq_app} we get
\beq
s^2 \tilde q_k(s) - q_k(0) s - \dot q_k(0) =  -\omega^2_k \tilde q_k(s) + c_k \mathcal{L}[g(x)]\,,
\eeq
where $\mathcal{L}[g(x)]$ is the Laplace transform of the composite function $g(x(t))$. Solving for $\tilde q_k(s)$ we obtain
\beq
\tilde q_k(s) = \frac{s \,q_k(0)}{s^2 + \omega_k^2} + \frac{\dot q_k(0)}{s^2 + \omega_k^2} + \frac{c_k \mathcal{L}[g(x)]}{s^2 + \omega_k^2}~,
\eeq
so that, coming back to the time domain,
\beq
q_k(t) = \frac{1}{2\pi i}\int_{0^+ -i\infty}^{0^+ + i\infty} \left[ \frac{s \,q_k(0)}{s^2 + \omega_k^2} + \frac{\dot q_k(0)}{s^2 + \omega_k^2}\right] e^{st}\, ds + \frac{c_k}{2\pi i}\int_{\sigma -i\infty}^{\sigma + i\infty} \frac{\mathcal{L}[g(x)]}{s^2 + \omega_k^2} e^{st}\, ds\,.
\eeq 
Recalling that, for $t > 0$,
\begin{subequations}
\begin{align}
\frac{1}{2\pi i}\int_{0^+ - i\infty}^{0^+ + i\infty} \frac{e^{st}}{s^2 + \omega_k^2}  ds & = \frac{\sin(\omega_k t)}{\omega_k}~,\\
\frac{1}{2\pi i}\int_{0^+ - i\infty}^{0^+ + i\infty} \frac{s\, e^{st}}{s^2 + \omega_k^2}  ds & = \cos(\omega_k t)\,,
\end{align}
\end{subequations}
we can write
\beq
q_k(t) = q^0_k(t) + \frac{c_k}{2\pi i}\int_{\sigma -i\infty}^{\sigma + i\infty} \frac{\mathcal{L}[g(x)]}{s^2 + \omega_k^2} e^{st}\, ds\,,
\label{eq:qk_invlapl}
\eeq
where
\beq
q^0_k(t) = q_k(0) \cos(\omega_k t) + \frac{\dot q_k(0)}{\omega_k} \sin(\omega_k t)\,.
\eeq
The second term on the r.h.s.\ of Eq.\ \eqref{eq:qk_invlapl} can be written as
\beq
\frac{c_k}{2\pi i}\int_{\sigma -i\infty}^{\sigma + i\infty} \frac{\mathcal{L}[g(x)]}{s^2 + \omega_k^2} e^{st}\, ds = \frac{c_k}{\omega^2_k}g(x(t)) - \frac{c_k}{2\pi i \omega^2_k} \frac{d}{dt} \int_{\sigma -i\infty}^{\sigma + i\infty} \mathcal{L}[g(x)]\frac{s}{s^2 + \omega_k^2} e^{st}\, ds \,,
\label{L_int}
\eeq
and since, using the convolution theorem, we can write 
\beq
\mathcal{L}[g(x)]\frac{s}{s^2 + \omega_k^2} = \mathcal{L}[g(x)] \mathcal{L}[\Theta(t) \cos(\omega_k t)] = \mathcal{L}[g(x) \ast \Theta(t)\cos(\omega_k t)]~,
\eeq
where $\Theta(x)$ is the Heaviside step function, substituting the above result into Eq.\ \eqref{L_int} and then back into Eq.\ \eqref{eq:qk_invlapl} we finally obtain
\beq
q_k(t) = q^0_k(t) + \frac{c_k}{\omega^2_k}g(x(t)) - \frac{c_k}{\omega^2_k} \frac{d}{dt}\int_0^t g[x(\tau)] \cos[\omega_k(t - \tau)]\, d\tau\,.
\eeq
\end{document}